\documentclass[aps,pre]{revtex4-1}
\usepackage[dvips]{graphics}
\usepackage{graphicx}
\usepackage{amsfonts}
\usepackage{amssymb}
\usepackage{amsmath}
\usepackage{subfigure}
\usepackage{color}

\begin{document}

\title{Vortex-Bright Soliton Dipoles: Bifurcations, 
Symmetry Breaking and Soliton Tunneling in a Vortex-Induced Double Well}

\author{M. Pola}
\affiliation{Zentrum f\"ur Optische Quantentechnologien,
  Universit\"at Hamburg, Luruper Chaussee 149, 22761 Hamburg, Germany}

\author{J. Stockhofe}
\affiliation{Zentrum f\"ur Optische Quantentechnologien,
  Universit\"at Hamburg, Luruper Chaussee 149, 22761 Hamburg, Germany}

\author{P. G. Kevrekidis}
\affiliation{Department of of Mathematics and Statistics, University of
Massachusetts,
Amherst, MA 01003-9305, USA}

\author{P. Schmelcher}
\affiliation{Zentrum f\"ur Optische Quantentechnologien,
  Universit\"at Hamburg, Luruper Chaussee 149, 22761 Hamburg, Germany}

\begin{abstract}
The emergence of vortex-bright soliton dipoles
in two-component Bose-Einstein condensates through bifurcations
from suitable eigenstates of the underlying linear system is examined. 
These dipoles can have their bright
solitary structures be in phase (symmetric) or out of phase (anti-symmetric). 
The dynamical
robustness of each of these two possibilities is considered and the
out-of-phase case is found to exhibit an intriguing symmetry-breaking
instability that can in turn lead to tunneling of the bright 
wavefunction between the two vortex ``wells''. We interpret this
phenomenon by virtue of a vortex-induced double well system, whose spontaneous
symmetry breaking leads to asymmetric vortex-bright dipoles, in addition
to the symmetric and anti-symmetric ones. The theoretical prediction of
these states is corroborated by detailed numerical computations.
\end{abstract}

\maketitle

\section{Introduction}

Within the booming field of Bose-Einstein condensation (BEC), the study
of coherent nonlinear states has its own considerable 
history~\cite{emergent,revnonlin,rab,djf}. The original explorations
were in the setting of repulsive interatomic interactions, and
especially so in the context of one-component settings, starting
over a decade ago~\cite{han1,nist,dutton,bpa}. These works
chiefly focused on the dark soliton i.e., the prototypical nonlinear state
therein, yet they were considerably hampered by instability effects
induced by the dimensionality of the system and/or the presence of
thermal effects. Nevertheless, more recent efforts using a variety
of techniques have been far more successful in generating robust
(dark) solitary wave states. Such techniques include 
phase-imprinting/density engineering \cite{hamburg,hambcol,technion}, 
matter-wave interference \cite{kip,kip2}, or dragging localized defects 
through the BECs \cite{engels}. 
The two-dimensional generalization of such (dark) states has as its
prototype the vortex waveform, which became possible~\cite{Matthews99} by 
using a phase-imprinting method between two hyperfine
spin states of a $^{87}$Rb BEC \cite{Williams99}.  Subsequent efforts
involved  the stirring of the BECs \cite{Madison00} above a certain critical 
angular speed
\cite{Recati01,Sinha01,corro,Madison01} which, in turn, 
led to the production of few vortices
\cite{Madison01} and even of very robust vortex lattices \cite{Raman}. Other
techniques including
dragging obstacles through the BEC \cite{kett99} or the nonlinear interference
of condensate fragments \cite{BPAPRL} have been also used for the production
of unit-charge vortices. 
Higher-charged vortex structures were produced \cite{S2Ket} and their dynamical
(in)stability has been examined. 

All of the above explorations were developed in the context of
one-component BECs. Yet, solitary wave states also exist in multi-component
BEC settings. In that context, of growing interest within the past few
years has been the study of dark-bright (DB) solitons that are supported in two-component \cite{BA} and even spinor \cite{DDB} condensates. These states can be 
thought of as ``symbiotic'', in that the bright
second component could not be sustained in the absence of the trapping dark
first component. Robust such states were first observed  
in the experiment of Ref.~\cite{hamburg} by means of a phase-imprinting 
method.

This, in turn, has led to experimental studies of numerous features of these
multi-component waves including the realization of DB 
soliton trains~\cite{engels1}, 
DB soliton oscillations and interactions~\cite{engels2,engels4}, 
as well as the possibility to create dark-dark breathing 
counterparts of these states (and multi-wave generalizations
thereof)~\cite{engels3,engels5}. In two dimensions, generalizations
of these states have been proposed in the form of vortex-bright (VB)
solitons, which were introduced about a decade ago
(see e.g.~\cite{skryabin} and references therein) 
and were recently further explored
in the work of~\cite{kody}. 

It is with the VB waveforms that we will concern ourselves
in the present study. Early experiments, such as the one of~\cite{anderson},
have illustrated the feasibility of realization of these
states. Additionally, recent studies of dynamical phenomena in
two-component condensates with considerable temporal and spatial
resolution and control~\cite{wieman,mertes,egorov} suggest that
the relevant coherent states can be explored further. Our aim here is
to explore this potential beyond the level of a single vortex-bright
soliton entity. In particular, recently in the one-component setting,
we have examined the bifurcation of few-vortex clusters 
(see~\cite{bifurcation} and references therein), most notably
the vortex dipole, but also the vortex tripole, quadrupole, vortex
polygons and larger scale crystals~\cite{stephan_physd}. It is then 
natural
to expect that similar bifurcations will arise in the two-component
setting and, in fact, that the bright solitons that are ``trapped''
within the vortex states will potentially bear different relative phases
(e.g. in phase or out of phase, as is the case with DB
solitons in~\cite{engels4}). It is relevant to mention here
that particular (in phase) realizations of such states have been
very recently proposed as realizable by means of numerical dragging 
experiments in the immiscible regime of the 
pseudo-spinor system in~\cite{gautam}. The excitations resulting
by dragging a laser beam through the two-component system had
earlier been explored in the miscible regime in~\cite{susanto}.

The vortex-bright dipoles, consisting of a vortex pair
of opposite circulation and a corresponding trapped bright soliton pair, 
will be the main theme herein.
We will start by providing the background of the relevant model and theoretical
setup, as well as a brief review of 
the properties of a single VB soliton in 
section II. Then, in section III, we will explain the different types of
[in phase (IP), or out of phase (OOP)] VB dipoles 
that exist and will present
their bifurcation from the (nonlinear) continuation
of underlying linear states in the form of dark-bright
soliton stripes. In section IV, we examine the stability of the VB
dipoles and recognize an instability of the OOP states and its
symmetry breaking and tunneling implications. These are then theoretically
explained in the form of an effective double well theory (as induced
by the vortices on the bright components) and its symmetry breaking
bifurcations, which gives rise to rather unexpected {\it genuinely asymmetric}
VB dipole states. We summarize our findings in section V, where we also present
some conclusions and possible directions for future studies.

 We should mention in passing that although our principal focus
herein will stem from BEC and atomic physics considerations,
relevant topics and ideas are, in principle, relevant for
nonlinear optics as well. In particular, structures
such as dark solitons~\cite{kivshar_luther} and 
vortices~\cite{yuripismen,dragomir} have been extensively studied
in the latter field. In fact, DB-soliton states were also first observed in 
optics experiments, where they were created in photorefractive 
crystals~\cite{seg1}, while their interactions were 
discussed in Ref.~\cite{seg2}. It is thus natural to expect that the
combination of multi-component and multi-dimensional settings therein
would also yield further potential for the realizability and 
observation of the coherent structures analyzed in the present work.

%

\section{Model Setup}

Our starting point will be the setting of 
quasi two-dimensional repulsive binary BECs, whose mean field description is 
given by the following set of equations:

\begin{align}
i \partial_{t} \psi_1(x,y,t) &= \left[ -\frac{1}{2} (\partial_x^2+\partial_y^2) + V(x,y) + g_1 |\psi_1|^2 + \sigma_{12} |\psi_2|^2 \right] \psi_1(x,y,t) \nonumber \\
i \partial_{t} \psi_2(x,y,t) &= \left[ -\frac{1}{2} (\partial_x^2+\partial_y^2) + V(x,y) + g_2 |\psi_2|^2 + \sigma_{12} |\psi_1|^2 \right] \psi_2(x,y,t) \text{.}
\label{th:eq: binGPE}
\end{align}
This  coupled dimensionless set of Gross-Pitaevskii  (GP) equations
describes the time evolution of the two components' order 
parameters $\psi_j$, $j \in \{1,2\}$. 
Time, length, energy and densities $|\psi_j|^2$ are measured in units 
of $\omega_z^{-1}$, $a_z$, $\hbar \omega_z$ and $(2\sqrt{2\pi}|a_{12}| a_z)^{-1}$, respectively;
$\omega_z$ and $a_z$ denote the oscillator frequency and length in the frozen $z$-direction, while $a_{11}$, $a_{22}$ and $a_{12}$ refer to the intra- and intercomponent scattering lengths. 
In the resulting dimensionless form of the equation, the coupling constants 
are $g_j = a_{jj}/|a_{12}|$, and $\sigma_{12}$ denotes the sign of $a_{12}$. 
From here, all equations will be
presented in dimensionless units and all the quantities which are plotted are dimensionless as well.

In the following, we will analyze 
the relevant case of binary condensates composed of ${}^{87}$Rb atoms in the two spin states ($F = 1$, $m_F = -1$) and ($F = 2$, $m_F = 1$), 
leading to dimensionless coupling constants of approximately
$g_1 = 1.03$, $g_2 = 0.97$, $\sigma_{12} = +1$~\cite{emergent}, 
and we will exclusively consider isotropic harmonic potentials 
$V(x,y) = \Omega^2(x^2 + y^2)/2 \equiv \Omega^2 r^2/2$, where $\Omega = \omega_r / \omega_z$ denotes the ratio of the in-plane and out-of-plane oscillator frequencies and is fixed to $0.2$. 
This trap strength is only selected for reasons of computational
convenience (and experimental realizability), but the phenomenology 
presented
below will not depend in any critical way on this selection for 
quasi-two-dimensional BECs.
Stationary solutions are obtained by factorizing $\psi_j(x,y,t) = \exp(-i\mu_jt)\phi_j(x,y)$, $j \in \{1,2\}$, where $\mu_1$ and $\mu_2$ are the two components' chemical potentials.

\begin{figure}[h!]
 \fbox{\includegraphics[width = 0.5 \textwidth]{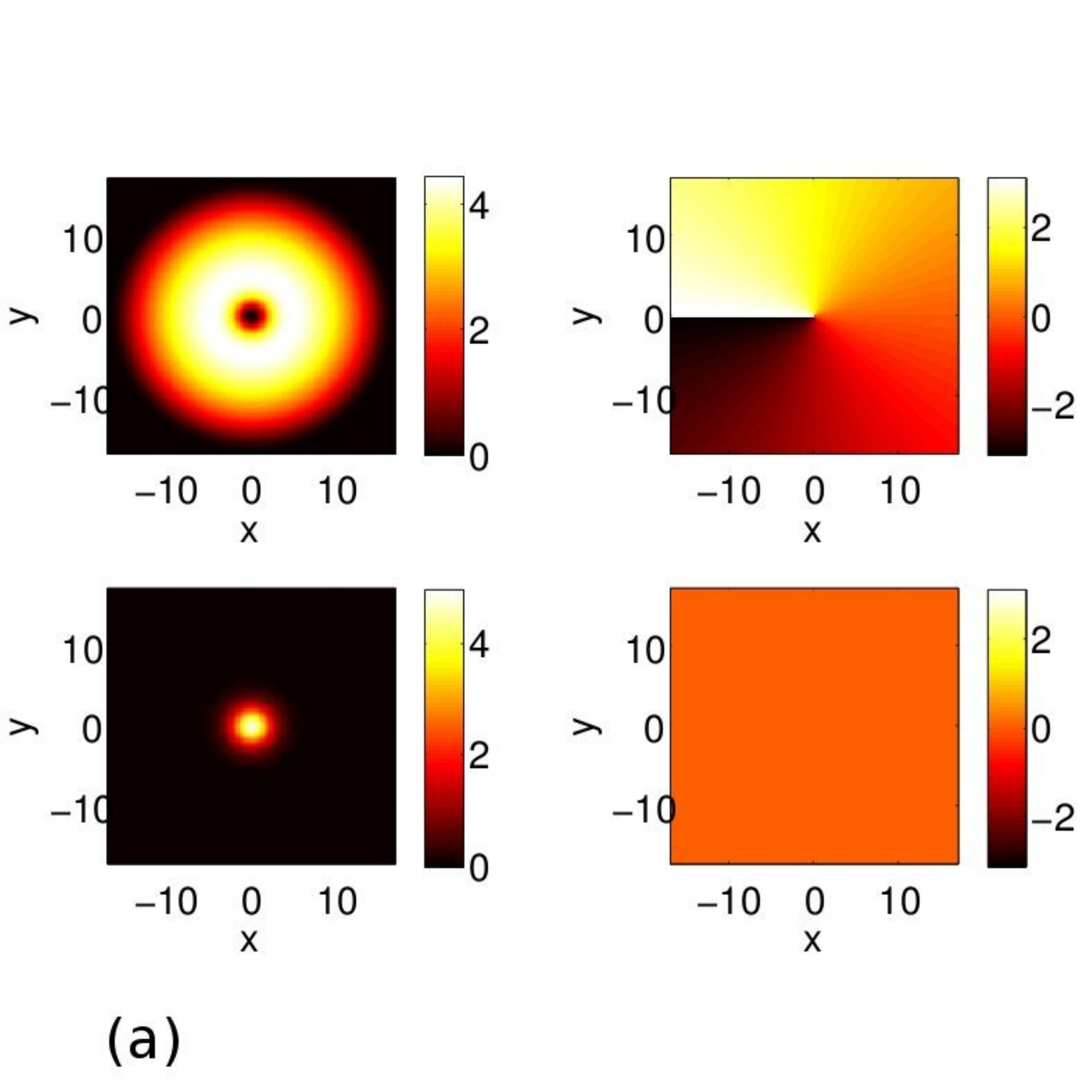}}
 \includegraphics[width = 0.4 \textwidth]{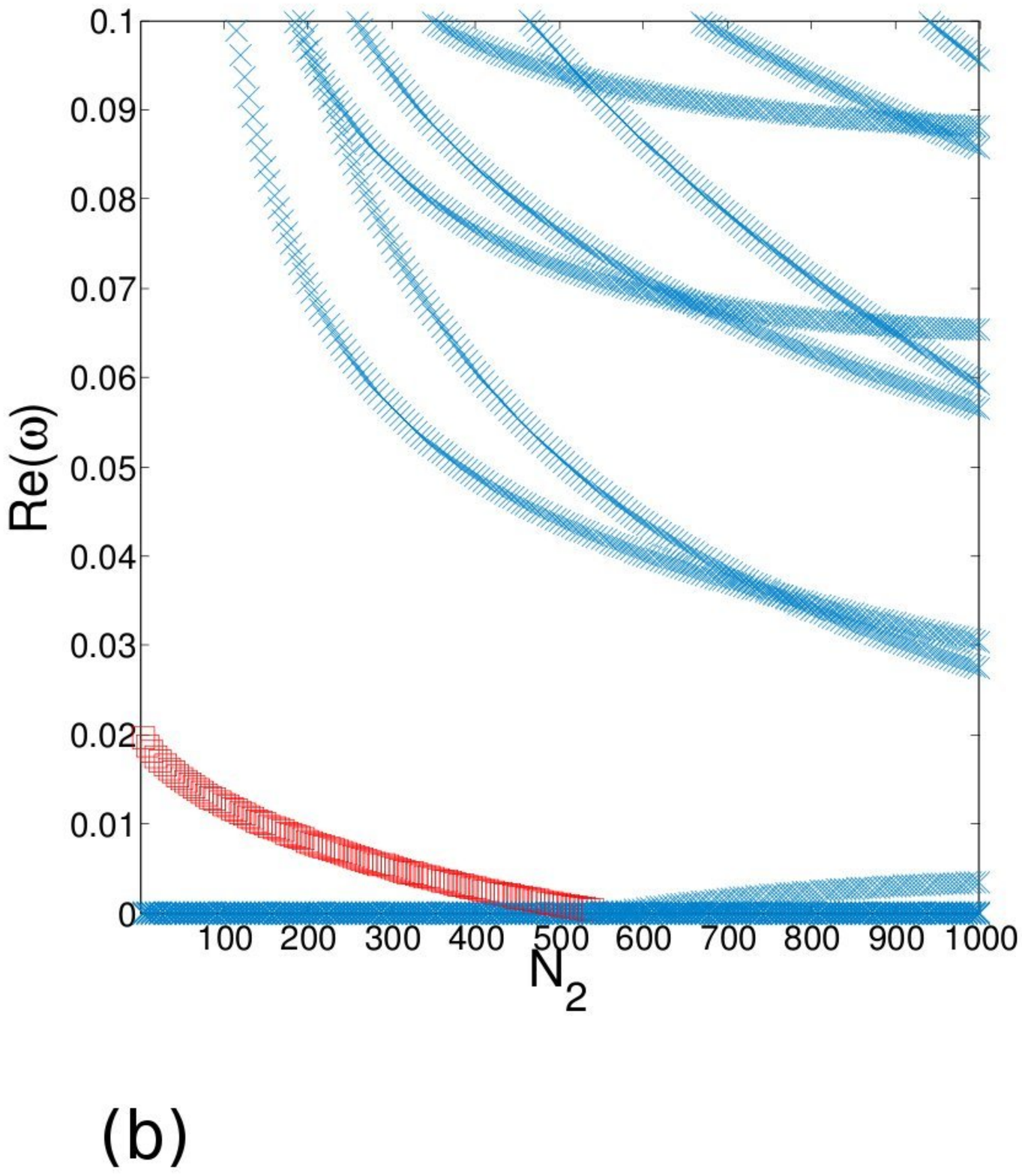}
 \caption{(Color online)  (a) Numerically calculated vortex-bright soliton solution to the two-component Gross-Pitaevskii equation obtained for $N_1 = 2000$ 
and $N_2 = 100$. The density (left) and the phase (right) of the two
components is shown. All the other state profiles shown throughout the present work will use the same color coding, with the colorbars ranging from zero to maximum density and $-\pi$ to $+\pi$, respectively.
 (b)  BdG spectrum of the vortex-bright soliton for $\mu_1=5.2$
as a function of $N_2$. Just the real part of the eigenfrequencies 
$\omega$ is shown, as the imaginary part is found
to be identically zero 
for all values of $N_2$.
Negative energy modes are indicated by darker square markers (red in the online
version).}
 \label{fig:vortexbright0}
 \end{figure}

The prototypical stationary solution of Eq.~\eqref{th:eq: binGPE} 
that will be the building block for our considerations 
is the single vortex-bright soliton state, whose density and phase profiles 
are shown 
in Fig.~\ref{fig:vortexbright0}. For such a state, the 
first component supports a 
vortex, which acts as an effective potential well for the second
component (of course, the alternative arrangement also exists where
the role of the components is interchanged). 
However, we will focus solely on the former case due to its 
unconditional stability; as discussed e.g. in~\cite{skryabin,kody}, in the
case where the components are interchanged, parametric regimes of instabilities
may arise. 

The linear stability (so-called Bogolyubov-de Gennes
or BdG) analysis is employed in order to consider the fate of
small amplitude perturbations and the potential robustness of the
solutions. This consists of imposing a perturbation to the stationary
solutions $\phi_j$ above in the form: 
\begin{eqnarray}
\psi_1(x,y,t)&=& \exp(-i\mu_1 t) \left\{\phi_1(x,y) + \varepsilon \left[a(x,y) e^{i \omega t}
+ b^{\ast}(x,y) e^{-i \omega^{\ast} t} \right] \right\},
\label{eq6}
\\
\psi_2(x,y,t)&=& \exp(-i\mu_2 t) \left\{\phi_2(x,y) + \varepsilon \left[c(x,y) e^{i \omega t}
+ d^{\ast}(x,y) e^{-i \omega^{\ast} t} \right] \right\}.
\label{eq7}
\end{eqnarray}
This leads at $\mathcal{O}(\varepsilon)$ (where $\varepsilon$ is a formal small
parameter) to an eigenvalue problem for the (in principle, complex)
frequency of excitations
$\omega$ and the corresponding eigenvector 
$[a(x,y), b(x,y), c(x,y), d(x,y)]^T$. Further details about the mathematical
structure of the BdG two-component problem can be found in~\cite{skryabin}.
For our purposes, it suffices to note that the Hamiltonian structure of
the resulting eigenvalue problem enforces that if $\omega$ is an eigenfrequency
so are $\omega^{\ast}$, $-\omega$ and $-\omega^{\ast}$. Hence, if 
eigenfrequencies with Im$(\omega) \neq 0$ exist, then the solution is
deemed to be dynamically unstable. 

A prototypical example of the BdG spectrum
of a VB solitary wave is provided in 
Fig.~\ref{fig:vortexbright0}(a). In this example, the spectrum 
is offered as a function of the (rescaled) number of atoms in the bright
soliton component $N_2 = \int \text{d}x\text{d}y |\psi_2|^2$, while the chemical potential of the dark (vortex)
component stays fixed at $\mu_1 = 5.2$. One important observation to 
make here is that there is a single negative energy mode~\cite{skryabin} in the
spectrum of the vortex for small (or vanishing) $N_2$. This mode
is illustrated by the square markers (which are red in the online version) in the relevant panel and is
well-known to correspond to the precession of the vortex within
the parabolic trap (see e.g.~\cite{bifurcation} and references
therein). However, it is noteworthy that this mode decreases in 
frequency due to the presence of the bright component and ultimately
crosses the zero frequency point. Yet, this crossing does {\it not}
produce an instability; the relevant  eigenfrequency pair remains real
but now the energy of the mode is positive signaling the transition of
the vortex state from a saddle in the energy landscape into a local
minimum thereof (this is a setting analogous to what is observed
for a vortex in the presence of rotation; see e.g. the relevant
discussion of~\cite{castin}). It is important to note that 
this observation is
in agreement with the results described in~\cite{anderson}, where it was 
observed that filled vortex cores exhibit slower precessions, and 
thus a decrease of the precession frequency is expected. 
It is relevant to note that similar observations have also been
obtained in the case of dark-bright solitons, originally through
the theoretical analysis of~\cite{BA} and have been experimentally
confirmed in the work of~\cite{engels2}.

\section{Vortex-Bright Soliton Dipole States}

\begin{figure}[h!]
 \fbox{\includegraphics[width = 0.4 \textwidth]{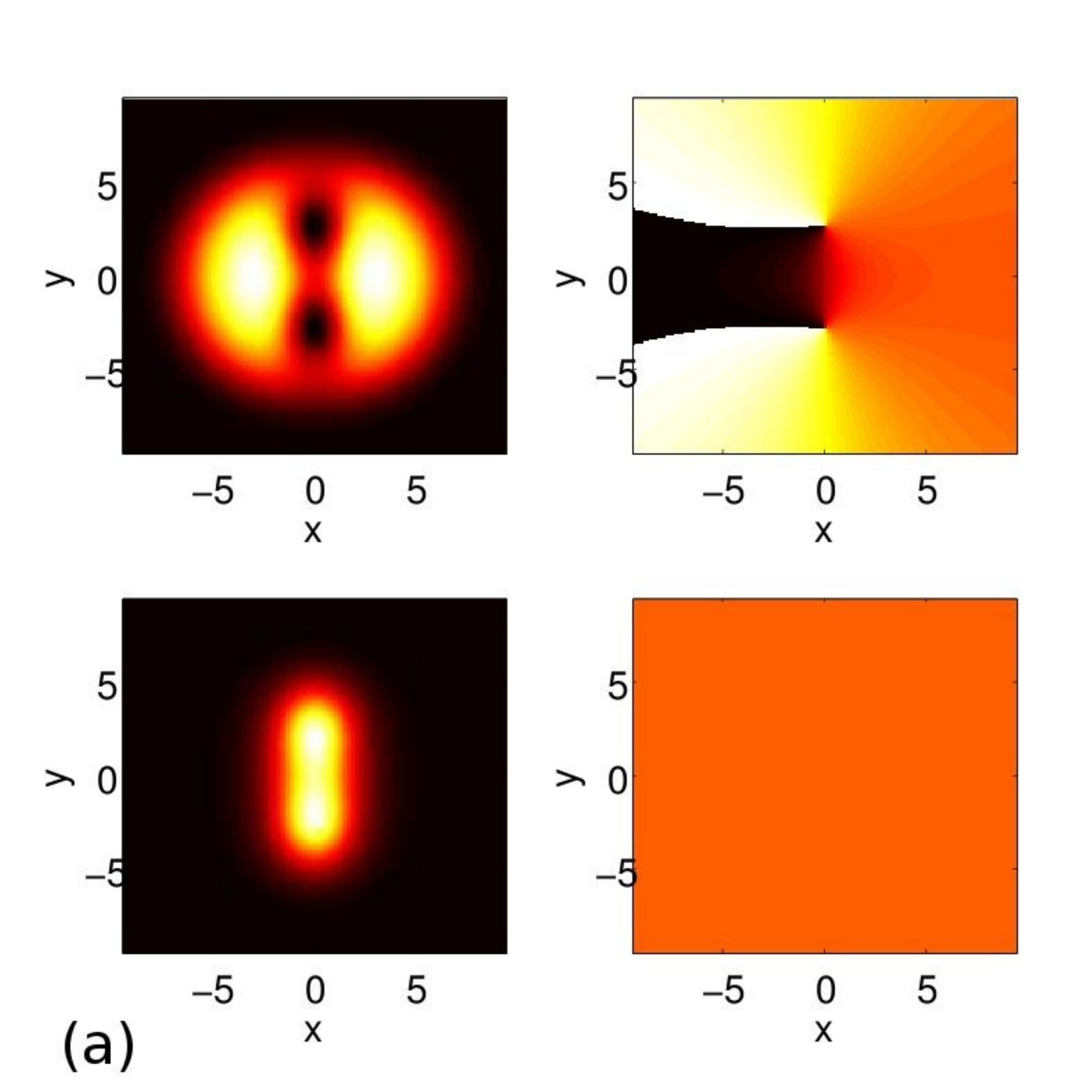}}
 \fbox{\includegraphics[width = 0.4 \textwidth]{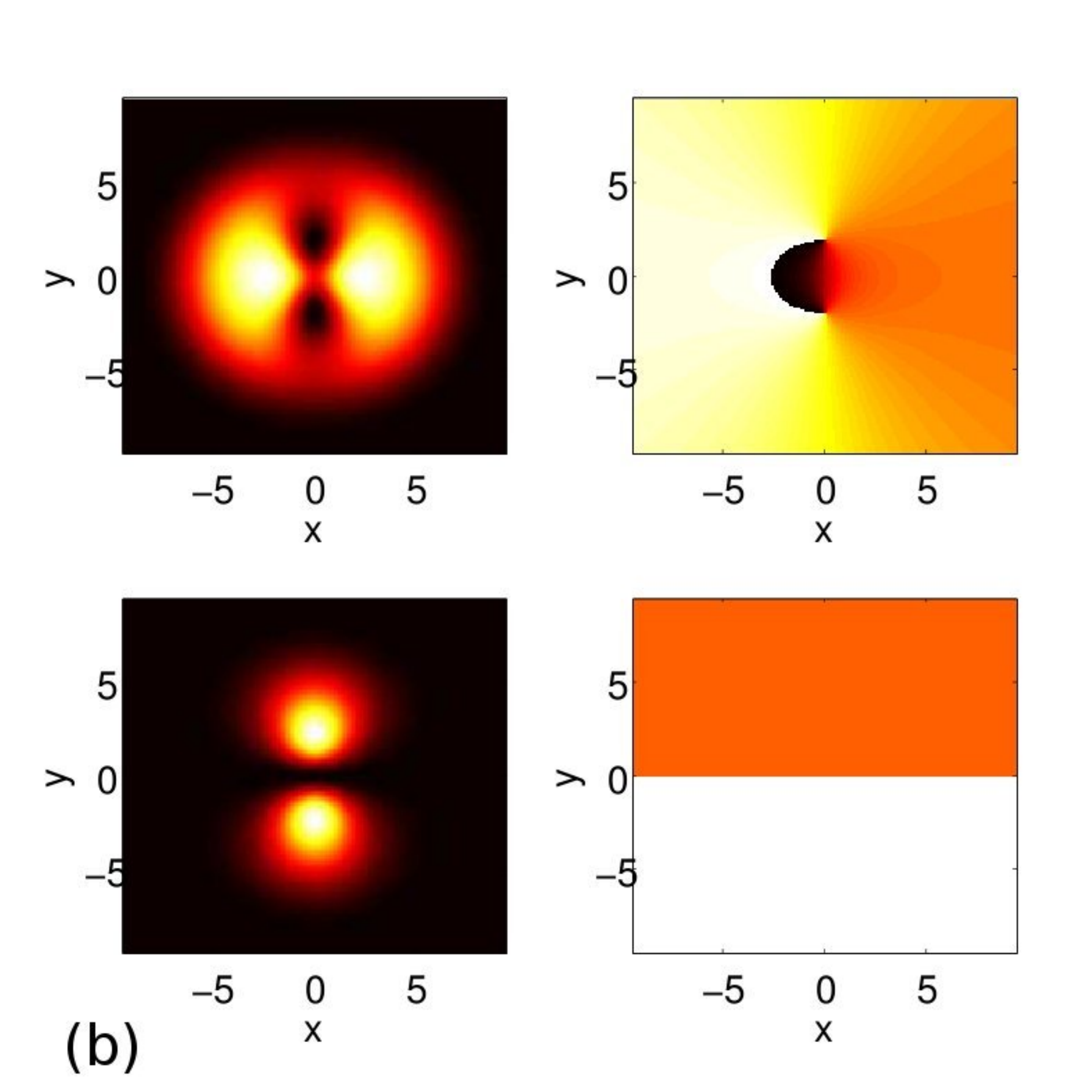}}
 \caption{(Color online) (a) Density (first column) and phase (second column)
of an in-phase vortex-bright dipole. (b) Density (third column)
and phase (fourth column) of the out-of-phase vortex-bright dipole.
Both examples are for $N_1=70$ and $N_2=20$.}
 \label{fig:vortexbright1}
 \end{figure}

Using the above fundamental building blocks, namely the single VB
solitons, we now look for bound states containing multiple such entities.
In the same spirit as in the work of~\cite{bifurcation} (see also 
e.g.~\cite{komineas} and references therein), the prototypical
relevant bound state is the VB dipole.
We have been able to identify two such states, which are both shown
in Fig.~\ref{fig:vortexbright1} for $N_1=70$ and $N_2=20$.
In both cases the first component contains 
two vortices located symmetrically with respect to the trap center, which are filled by the 
second component. The difference between the two dipoles is evident in the 
phase profile: the two bright solitons filling the vortex 
cores can have either the same phase, or there may 
be a phase difference of $\pi$ between them. 

\begin{figure}[h!]
 \fbox{\includegraphics[width = 0.28 \textwidth]{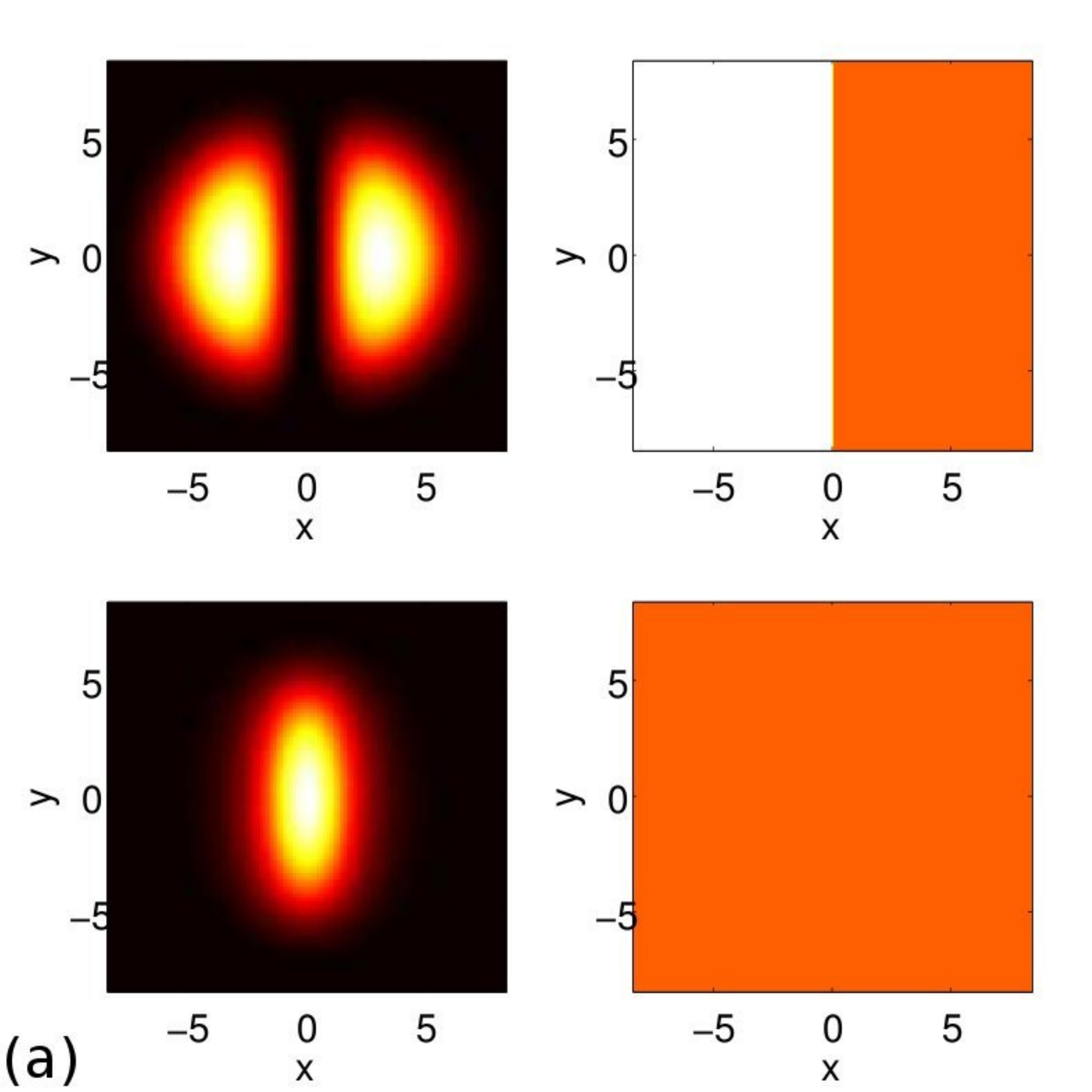}}
 \fbox{\includegraphics[width = 0.28 \textwidth]{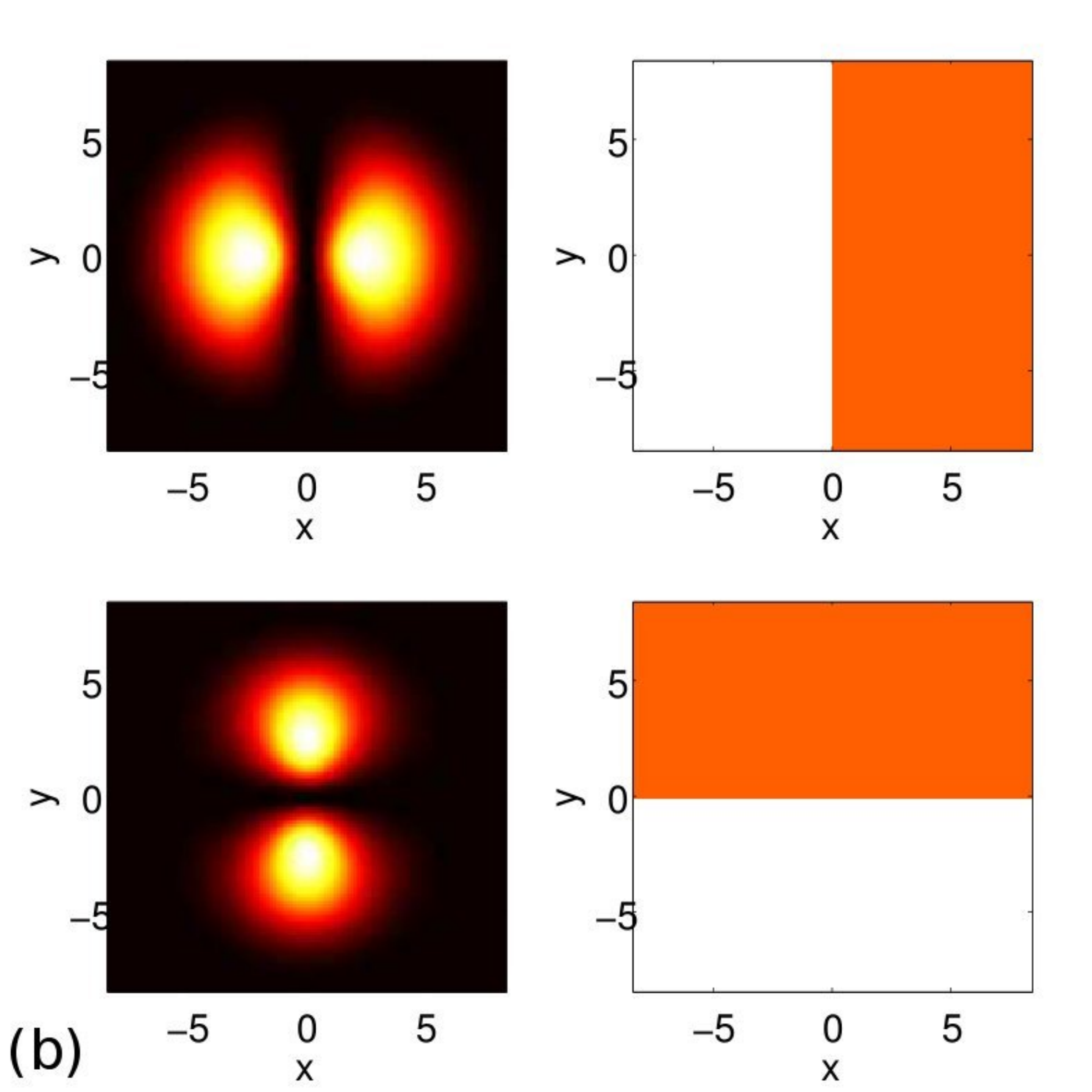}}\\
\includegraphics[width = 0.35 \textwidth]{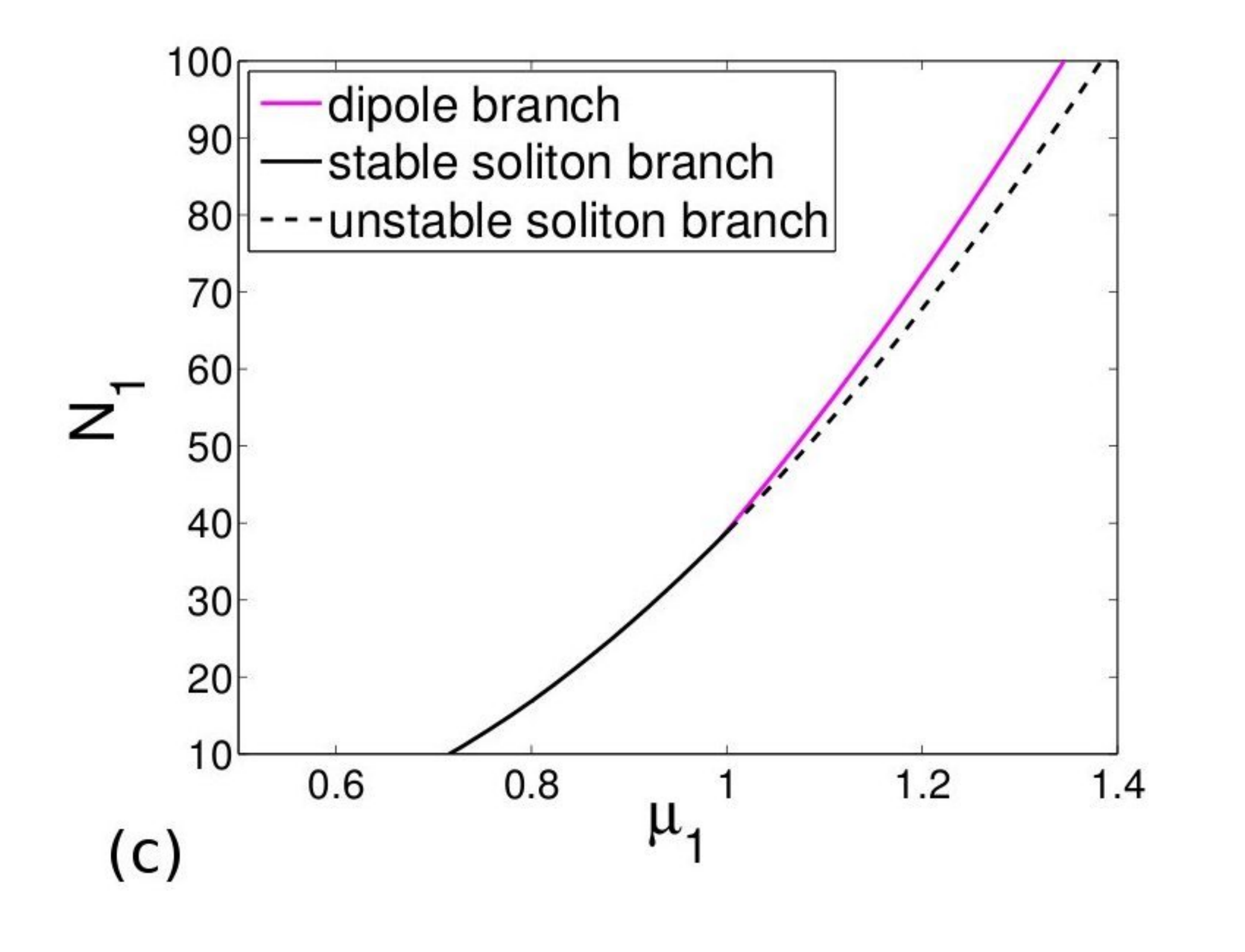}
  \includegraphics[width = 0.35 \textwidth]{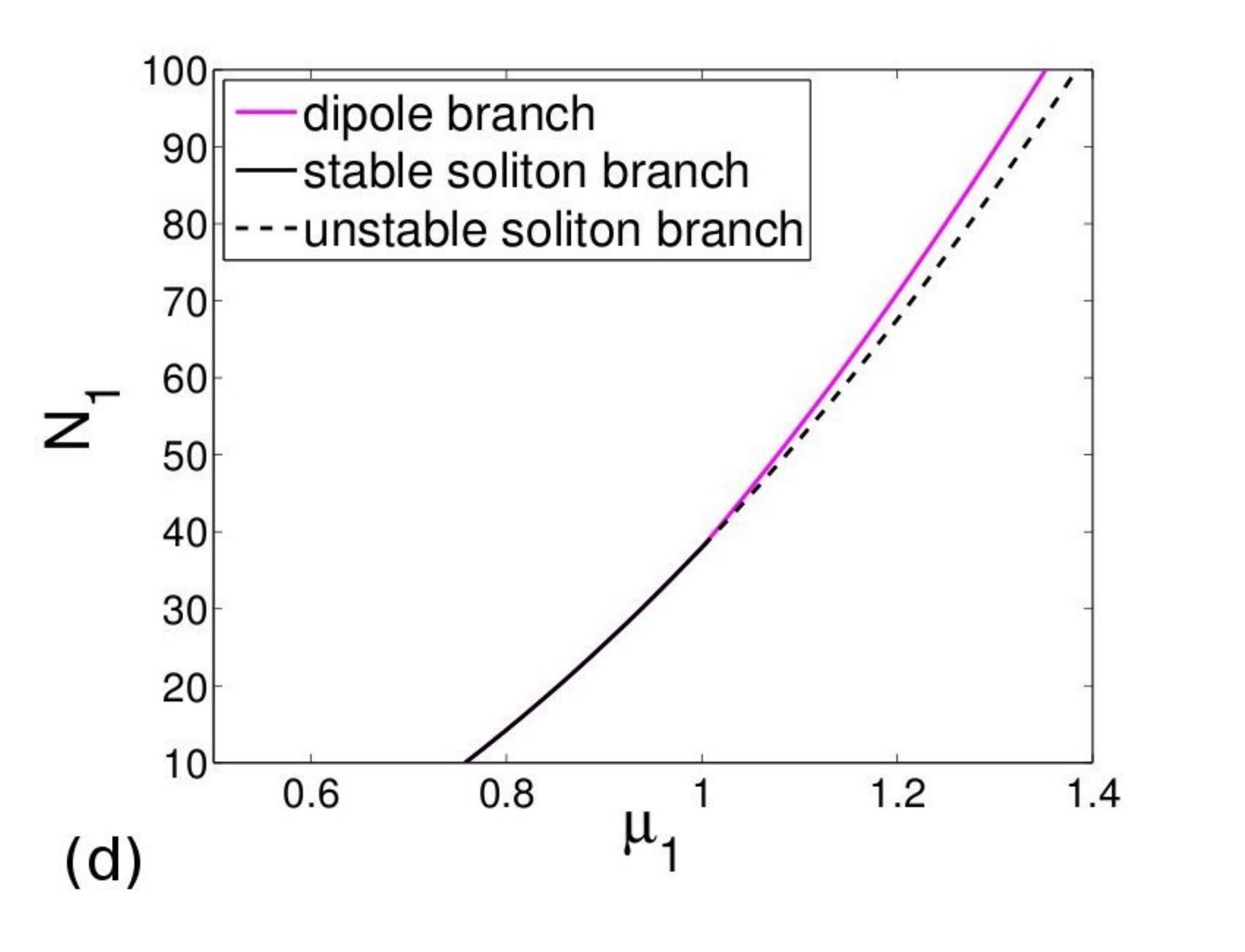}\\
\includegraphics[width = 0.27 \textwidth]{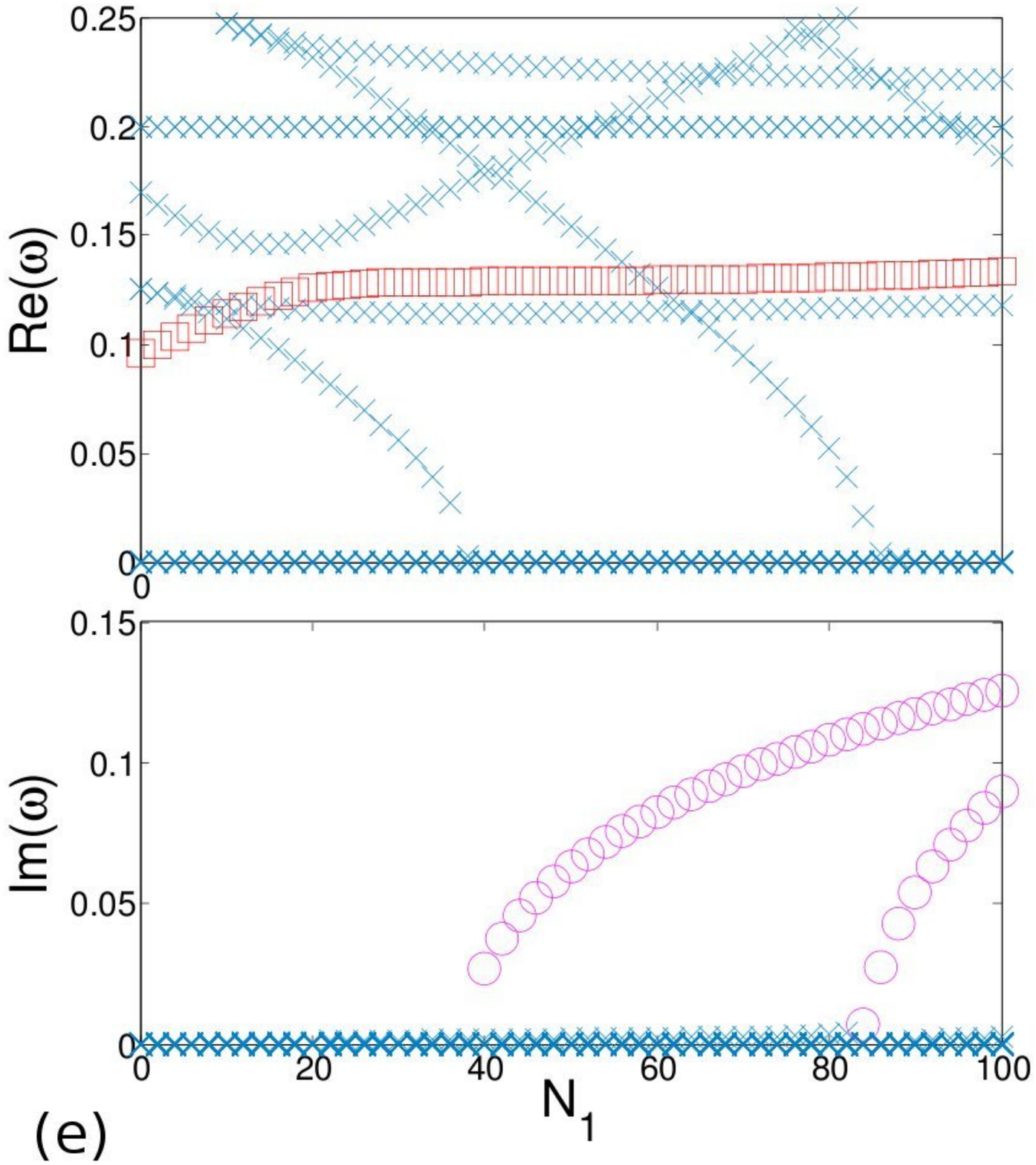}
\vspace{-5mm}
 \includegraphics[width = 0.27 \textwidth]{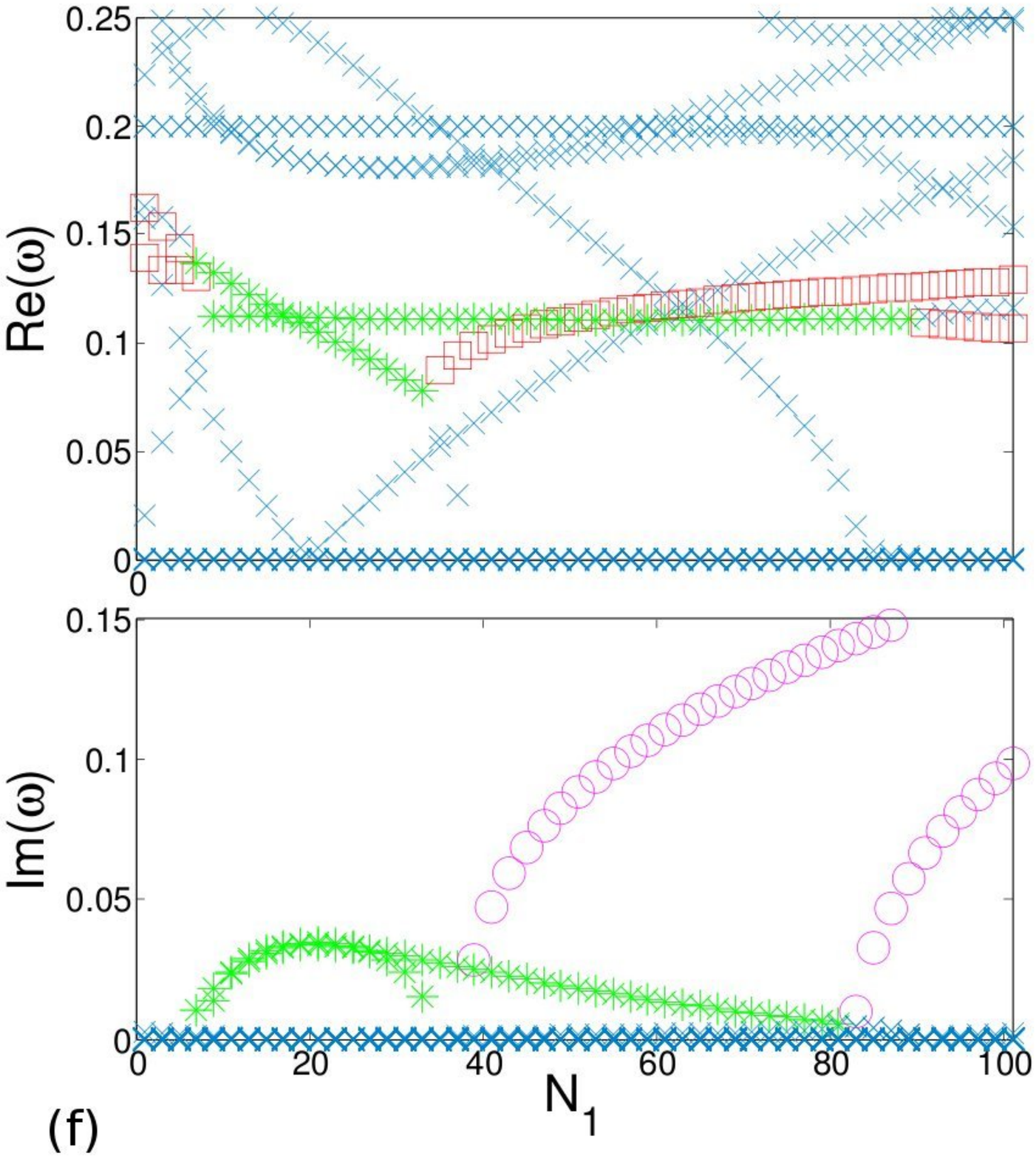}\\
\vspace{5mm}
 \includegraphics[width = 0.7 \textwidth]{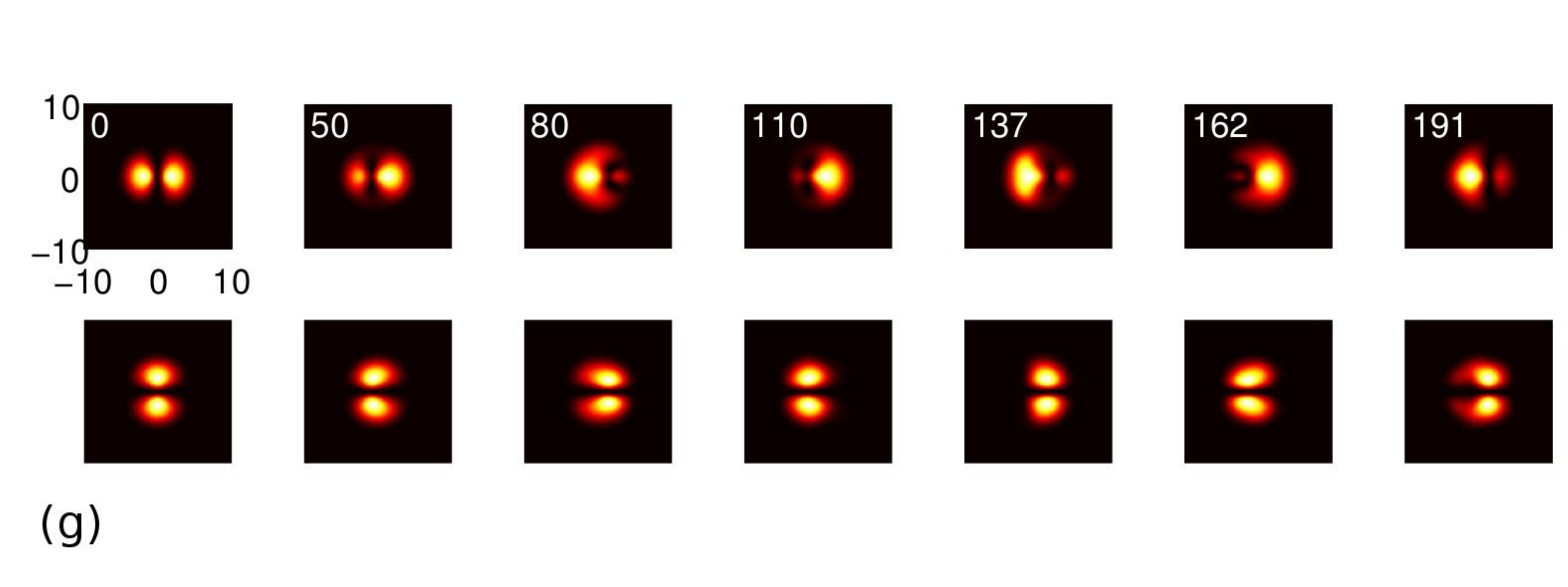}\vspace{-5mm}
 \caption{(Color online) (a) Density and phase of the in-phase  dark-bright
soliton stripe,  $N_1=38$, $N_2=20$. (b) Density and phase of the out-of-phase dark-bright
soliton stripe, $N_1=38$, $N_2=20$. In the second row we see the bifurcation of the IP and OOP ((c) and (d), respectively) vortex-bright soliton
dipoles branch from the
corresponding parent states (whose density and phase profiles are depicted in (a) and (b)). 
The third row's panels contain the BdG analysis of the parent states (in (e) and (f) we see the IP and the OOP case, respectively) of the top
panels, illustrating the instability (through an imaginary eigenfrequency, signaled by pink circle markers)
that is induced by the emergence of the vortex-bright dipoles. In addition,
for the out-of-phase case, the lines formed by ``$\ast$`` (green in the online version) markers correspond to the existence of oscillatory instabilities through
complex eigenfrequency quartets. In this case there are two anomalous
modes (depicted by red square markers), while there is only one such for the in phase stripe. (g) Time evolution of the out-of-phase dark-bright soliton stripe after being perturbed with the eigenvector of one of the oscillatorily unstable modes 
 at $N_1 = N_2 = 20$. The top and bottom 
lines depict the density profiles of the dark and bright component, respectively, evaluated at seven moments of the simulated evolution. 
The elapsed (dimensionless) time is indicated in the panels' upper left corners. }
 \label{fig:vortexbright2}
 \end{figure}

The emergence of the dipole state branches can be thought of in terms of a bifurcation picture in an equivalent way to the one described for 
one-component BECs in \cite{bifurcation}. The two possible dipole branches, 
distinguished from the presence 
or absence of the phase jump of $\pi$ in the bright component, 
arise at critical values of $N_1$ from two states, which can be regarded 
as generalizations of the dark soliton stripe observed in one-component 
condensates~\cite{djf}.  The density and phase profiles of 
these two solitonic states  are shown in the top panels of Figs.~\ref{fig:vortexbright2}(a) and ~\ref{fig:vortexbright2}(b), 
where it is evident that the two states 
are again distinguished by the presence or absence of a $\pi$-phase 
jump in the bright component. 

In the first case of the IP VB dipole, the parent state can be
seen to consist of a dark soliton stripe in the first component
accompanied by a bright stripe in the second component, which
is trapped inside the dark stripe. In the notation of the
corresponding linear limit of the two single-particle
Hamiltonians, the relevant state of the first component is
the $|1,0\rangle$ state (i.e., the first excited state in the
x-direction) while that of the second component
is the $|0,0\rangle$ ground state of the latter. More generally,
the symbolism $|m,n\rangle$ is used to denote the states
$H_m(\sqrt{\Omega}x) H_n(\sqrt{\Omega}y) e^{- \Omega r^2/2}$ where $H_m$ stands for the
\textit{m}-th Hermite polynomial and normalization constants have been omitted. 
These are stationary states of the underlying linear problem with eigenvalue $\mu=\Omega (m+n+1)$, 
from which the soliton stripe solutions arise in the presence of the effective nonlinearity induced by interatomic
interactions. It is the nonlinearity and immiscibility of the two
components that leads to the $|0,0\rangle$ state of the second component
being elongated (as opposed to circularly symmetric) in the
dark-bright stripe of Fig.~\ref{fig:vortexbright2}(a).
Corresponding states in a ring (as well as in diagonal stripes) form
have recently been addressed in~\cite{jan2}. In the case of the OOP
VB dipole, shown in  Fig.~\ref{fig:vortexbright2}(b), the parent state is the $|1,0\rangle$  of the first
component coupled to the $|0,1\rangle$ of the second component.

While the bright component remains roughly invariant as the transition point
is crossed, the dark component develops two singular points in the 
phase corresponding to the density vanishings and the opposite circulation
characteristic of a vortex dipole~\cite{bifurcation}.
The relevant bifurcations occur approximately at the same critical 
$N_1$ ($N_{1}^{cr} \approx 40$, with $N_2$ fixed at $20$). This is explicitly 
depicted in Figs.~\ref{fig:vortexbright2}(c) and ~\ref{fig:vortexbright2}(d) for
the two different ``stripe'' states, 
where the $N_1(\mu_1)$ bifurcation diagrams are shown. 
These approximately equal critical 
values of $N_1$ can, arguably, be expected 
due to the small number of particles of the second component, a statement
which becomes exact in the limit of $N_2 \rightarrow 0$.

The BdG analysis of the two stripe states is shown in Figs.~\ref{fig:vortexbright2}(e) and~\ref{fig:vortexbright2}(f) and allows to relate the emergence of the dipole 
state to the changing of stability of the parental branch (the IP and OOP solitonic states).

From the BdG spectra, in fact, one observes the
 emergence of an imaginary eigenfrequency pair for 
$N_{1}^{cr} \approx 40$, which is the value of $N_1$ at which the bifurcation actually occurs. 
Imaginary modes are here depicted by circle (pink in the online version) markers.
The "$\ast$" (green in the online version) markers are used to identify complex modes which are related
 to oscillatory instabilities, resulting from the collision
of positive energy modes (associated with the ``background''
on top of which the coherent solitonic structures exist) with
negative energy modes that are associated with the solitonic
structures (and the fact that they are not ground states
of the system). For detailed discussions of the such instabilities
and their origin,
 see ~\cite{emergent}, ~\cite{revnonlin} and ~\cite{skryabin}. 

The complex nature of such eigenvalues (or eigenfrequencies)
leads to a part associated with oscillation (connected to the
real part of such eigenfrequencies) and a part associated with 
growth (the imaginary part of such eigenfrequencies).
Hence, what one should expect upon perturbation of
such eigenmodes is an oscillatory growth that should lead
to a destabilization and eventual destruction of the relevant
solitonic structures (of Figs.~\ref{fig:vortexbright2}(a) and~\ref{fig:vortexbright2}(b)).
This effect is evident in Fig.~\ref{fig:vortexbright2}(g), where the propagation of the OOP soliton for $N_1=20$ and $N_2=20$, after the excitation of one of the 
two complex eigenmodes, is shown. Exciting the other mode, one obtains an equivalent evolution of the system, by interchanging the roles of the two components: the oscillation is 
observed in the other component. 
As is evident from the BdG spectra, here an oscillatory instability exists for the OOP solitonic state but not for the IP solitonic state. 
In this context, let us point out that the IP stripe bears only one
negative energy mode, while a second such anomalous mode is present in the OOP state's spectrum.
As a result, the IP configuration is less prone to oscillatory
instabilities emerging from collisions of anomalous and background modes than the OOP one.
The presence of one anomalous mode for the IP state and of two such modes for the OOP state is intuitively expected by the out-of-phase or excited
state nature that the former bears only in the first component,
while the latter has that type of structure in both components.
I.e., in the second component, the former configuration features
a ground state, symmetric waveform, while the latter has an 
anti-symmetric first excited state waveform. This type
of characteristics will be of critical relevance to the
considerations that follow below, regarding tunneling effects
and symmetry-breaking bifurcations.

\clearpage

\section{Tunneling Dynamics and Symmetry-Breaking Bifurcations}

\begin{figure}[h!]
 \centering
 \includegraphics[width = 0.3 \textwidth]{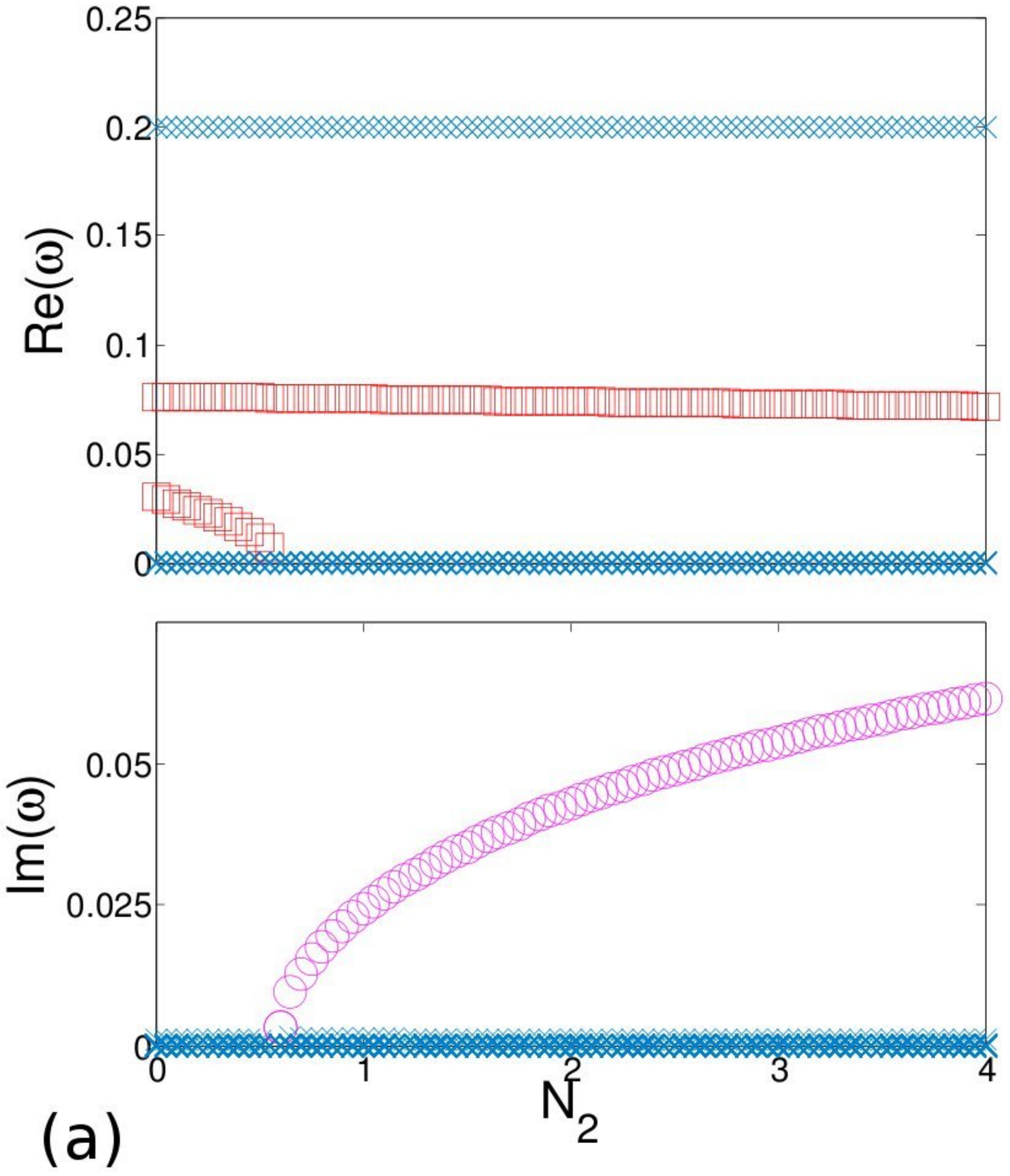}
 \includegraphics[width = 0.9 \textwidth]{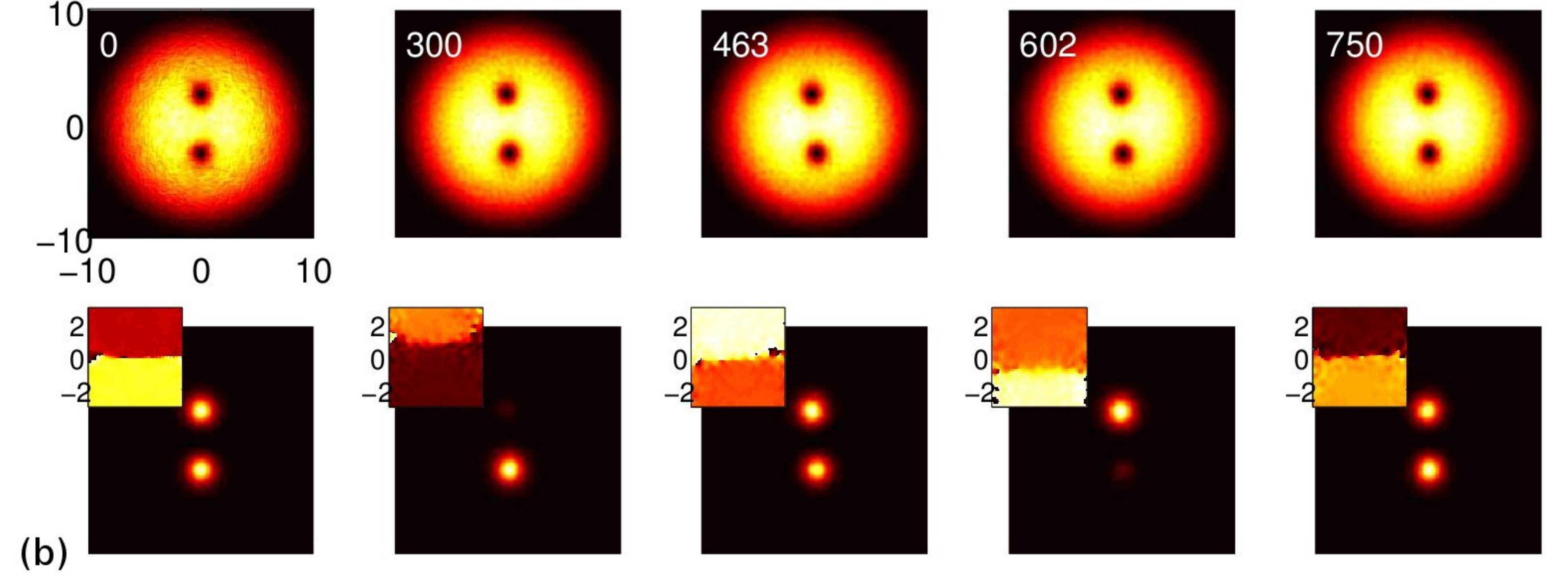}
  \caption{(Color online) (a) BdG spectrum of an out-of-phase vortex-bright soliton dipole for $N_1=220$ as a function of $N_2$. According to our convention, 
pink circle markers are related to unstable imaginary modes, while red square markers identify anomalous modes.
(b) Time propagation of the same state at $N_1 = 220$ and $N_2 =  1$. Again, the top and bottom 
lines depict the density profiles of the dark and bright component, respectively. At the upper left corners of the pictures of the second row the phase of the bright component, 
evaluated at the same time steps as the density profiles, is shown. The substantial addition of white noise (that seeds the instability) can be
seen in the pixelization of the corresponding dynamical panels.}
 \label{fig:vortexbright3}
 \end{figure}

Let us now come back to the vortex-bright soliton dipoles and especially
their dynamical stability through the BdG analysis. The in-phase VB dipole 
appears to be stable for arbitrary 
values of $N_1$ and $N_2$, while the out-of-phase VB dipole shows purely 
imaginary modes arising at certain values of $N_1$ and $N_2$ 
(when the state is scanned over $N_1$ or $N_2$, respectively), see Fig.~\ref{fig:vortexbright3}(a). 
Exciting this unstable mode for the OOP dipole with $N_1 = 220$ and $N_2 = 1$ by adding some white noise and 
letting the perturbed states propagate in time, the evolution
shown in Fig.~\ref{fig:vortexbright3}(b) is obtained.

The time evolution of the density profiles shows the size of one of the bright solitons increasing while the other one is getting smaller, 
then the opposite situation is observed and this trend with enlarging and shrinking bright solitons is repeating in the course of the propagation. 
After an initial transient phase during which the bright component's asymmetry builds up (first to second time step shown in Fig.~\ref{fig:vortexbright3}(b)), we find periodic oscillations. 
This observed time evolution has an immediate physical interpretation as concerns the bright component. A fraction of the corresponding 
particles within the second component move from 
one vortex core to the other during the time propagation of the system.
This can be explained by considering the dark component's density 
as acting in the form of 
an effective potential for particles of the bright one. 
I.e., the particles of the second component can be considered as 
tunneling within the vortex-induced (i.e., formed
by the vortex cores) double well potential. Within this very potential,
this suggests the possibility of a spontaneous symmetry breaking bifurcation,
as responsible for the observed phenomenology.

In order to validate this assumption, we consider the analytical model, 
developed in~\cite{symmetry}, based on a two-mode expansion, 
which is used to determine stationary 
states and to study the symmetry-breaking bifurcations occuring in 
one-dimensional, single-component repulsive BECs 
confined in double-well potentials. The model predicts that for a two-mode
decomposition of the equation involving a symmetric ground
state and an anti-symmetric first excited state, there exists
a symmetry breaking bifurcation both for attractive and for
repulsive interactions. The bifurcation emerges from (and
destabilizes) the anti-symmetric,
first-excited state in our repulsive interaction case (while it
stems from and destabilizes the symmetric ground state in the
attractive interaction case). This is consonant with the principal
observation above of the generic stability of the IP VB, which has
a symmetric second component and the symmetry breaking associated
destabilization of the OOP VB with the anti-symmetric second
component.

The key quantitative observation now is that if we freeze
the first component (assuming that it forms the vortex-core double
well potential), then the theory of~\cite{symmetry} can be applied
directly but for the effective double well potential
in the form
$V_\text{eff}(x,y) = V(x,y) +  |\psi_1(x,y)|^2$, 
where $V(x,y)$ is the trapping harmonic potential.
The equation for the second component can then be rewritten as 
\begin{equation}
i \partial_{t} \psi_2(x,y,t) = \left[-\frac{1}{2} \Delta_{2D} + V_\text{eff}(x,y) + g_2 |\psi_2|^2\right] \psi_2(x,y,t),  
\label{th:eq: binGPEcomp2a}
\end{equation}
which upon the rescaling $ \tilde \psi_2(x,y) = \sqrt{g_2} \psi_2(x,y)$ leads
to a standard single-component (two-dimensional) double well setting.
It should be noted that the potential of the double well  
clearly depends on $N_1$ (through its dependence on $ |\psi_1(x,y)|^2$).
For each $N_1$ (in multiples of $5$) in the interval 
$[0,300]$ the density profile of the 
dark component of the VB dipole state 
with $N_2 = 0$ has been utilized (to form the corresponding effective 
potential). For each considered value of $N_1$, one can diagonalize the 
obtained Hamiltonian $H =-\frac{1}{2} \Delta_{2D} + V_\text{eff}$,
and keep the two lowest 
energy eigenstates, namely the symmetric ground state and the 
anti-symmetric first excited state of the single particle
operator with the effective potential. These, denoted hereafter
as $u_0(x,y)$ and $u_1(x,y)$, will be used for the two
mode reduction
in the spirit of~\cite{symmetry}.

The fundamental analytical prediction of the two-mode approximation is
that as $N_2$ (and $\mu_2$) is increased, a critical point will be reached,
given by

\begin{equation}
 N_{2cr} = \frac{\Delta \omega}{(3B - A_1) g_2},
\label{th:eq:Nc2}
\end{equation}

whereafter the anti-symmetric branch (interpreted here as the OOP VB dipole)
will become unstable. Past this critical point, an asymmetric branch
(i.e., an asymmetric VB dipole) should emerge as a stable configuration. 
In the expression of Eq.~(\ref{th:eq:Nc2})
$A_1 = \int u_1^4 \text{d}x \text{d}y$ and 
$B = \int u_0^2 u_1^2 \text{d}x \text{d}y$ are overlap integrals of the two lowest energy eigenstates of the linear Schr\"odinger problem 
and $\Delta \omega = \omega_1 - \omega_0$, where $\omega_0$ and $\omega_1$ 
are the energy 
eigenvalues corresponding to $u_0$ and $u_1$.
The presence of the factor $g_2$ in the denominator ensures consistency 
with the scaling of $\psi_2$ discussed above.

The critical chemical potential is calculated from the 
particle number by making use of the 
expression $\mu_{2cr} = \omega_0 + 3 B N_{2cr}$~\cite{symmetry}.

\begin{figure}[h!]
\includegraphics[width = 0.35 \textwidth]{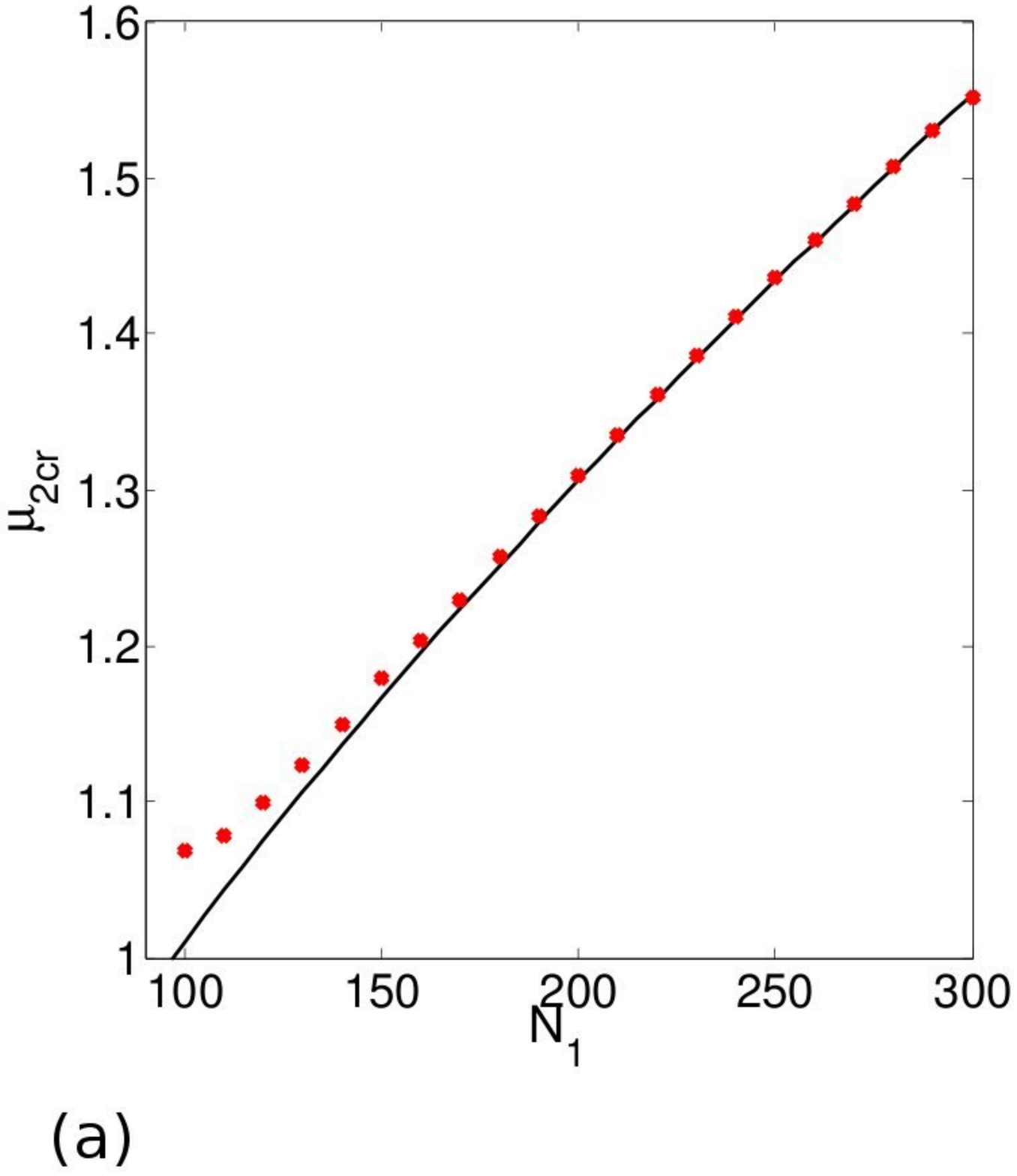}
\includegraphics[width = 0.35 \textwidth]{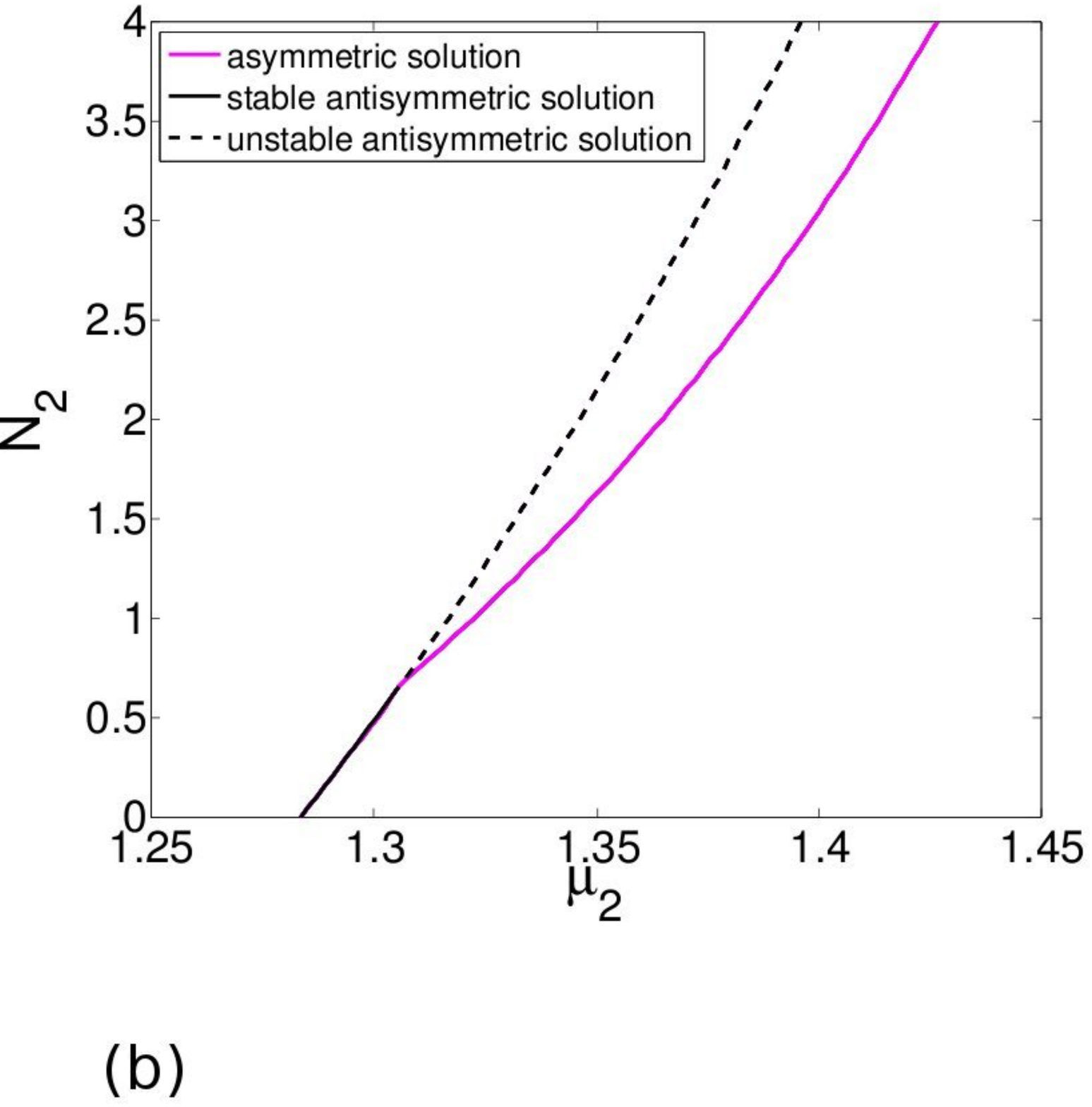}
\fbox{\includegraphics[width = 0.27 \textwidth]{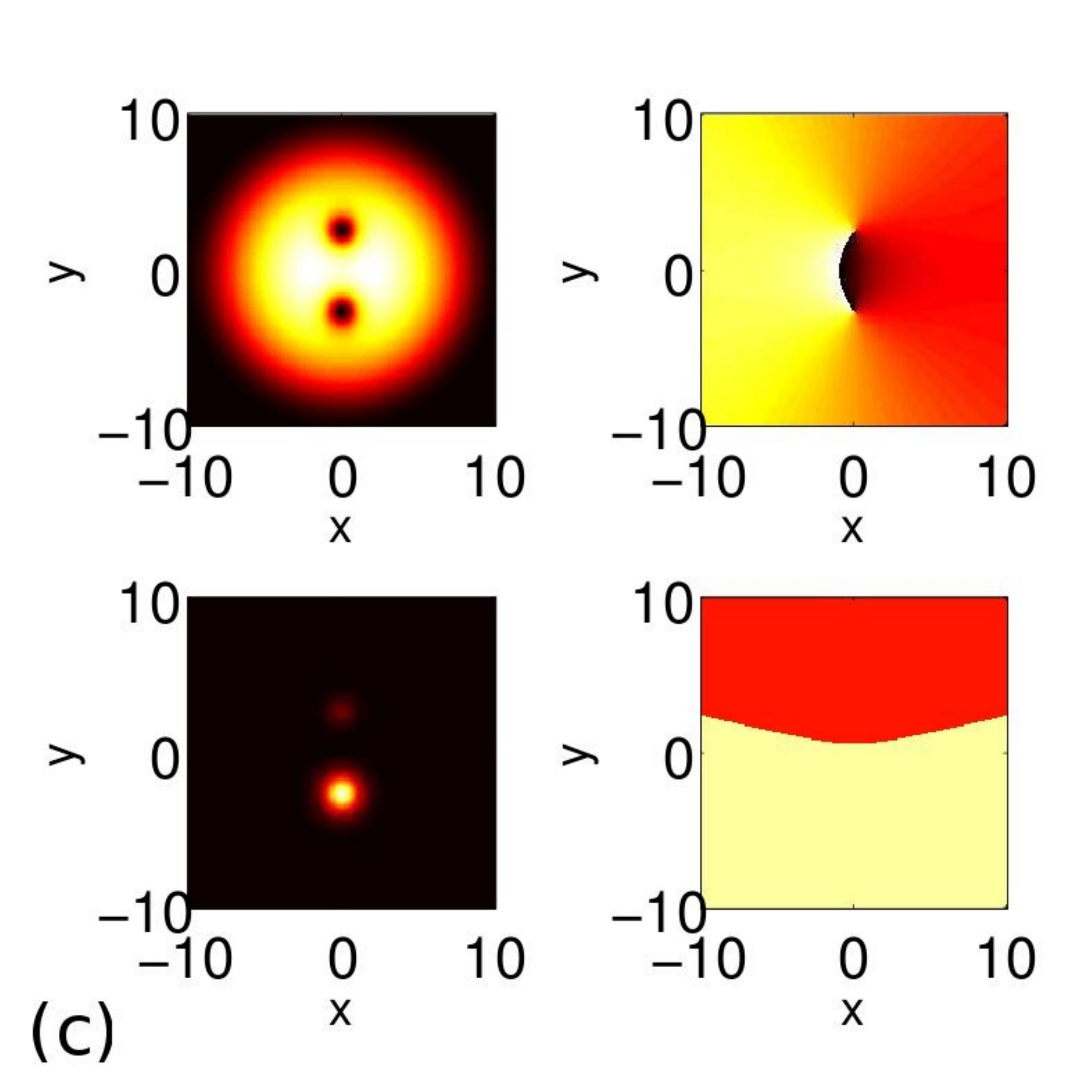}}
\caption{(Color online) (a) The critical points for the bifurcations of asymmetric states from the OOP dipole ($\mu_{2cr}$ for different
values of $N_1$) are depicted by red circle points and are compared
to the corresponding values of the theoretical two-mode
prediction (solid black line). (b) The bifurcation diagram demonstrates the existence
of an asymmetric vortex-bright dipole, as bifurcating from the
out-of-phase one. (c) Example of
density and phase profiles of such a symmetry-broken state.}
 \label{fig:vortexbright5}
 \end{figure}

Repeating this procedure for each considered $N_1$,
we get the critical 
values of $\mu_2$, which are compared with the numerically 
obtained data; see Fig.~\ref{fig:vortexbright5}(a).
These latter are 
obtained by making a scan within a suitable interval of $N_2$ of the 
OOP VB dipole, and then performing the BdG analysis of the 
resulting branch of solutions. In the ensuing BdG spectrum, the bifurcation point can be identified as the onset of instability of the anti-symmetric VB dipole.
For a relevant example, see again Fig.~\ref{fig:vortexbright3}(a), for the scan obtained for $N_1 = 220$, 
where we can see the imaginary mode emerging at $N_2 \approx 0.6$ (generally,
the relevant critical values $N_2^{cr}$ are of order unity). At this critical value of $N_2$, for which the antisymmetric dipole becomes 
unstable (or, equivalently, at the corresponding critical value of $\mu_2$), we expect the bifurcation of a stable asymmetric state, which has been actually verified, as will be shown below.

From Fig.~\ref{fig:vortexbright5}(a) it is evident that the 
agreement between analytical predictions and numerical 
results is very good for what concerns $\mu_{2cr}$ for $N_1 \gtrsim 120$. For lower values of $N_1$ numerics and analytical predictions 
start to be in disagreement with each other and this has the following explanation: 
by decreasing $N_1$, the height of barrier between the two wells forming the effective potential 
$V_\text{eff}(x,y) = V(x,y) + |\psi_1(x,y)|^2$, 
is also decreased. Thus, for low $N_1$, the density $|\psi_1(x,y)|^2$ provides a much 
lower contribution to the potential, and the effect of the harmonic trap is much more appreciable. This influences the form of the energy 
spectrum: the gap separating the almost equal lowest energy eigenvalues $\omega_0$ and $\omega_1$ from the larger eigenvalues $\omega_2$, 
$\omega_3$..., which is typically large for a high barrier, if the latter is lowered becomes smaller and smaller, 
until the spectrum of the harmonic oscillator, 
formed by equidistant eigenvalues, is reached. But, if this is the case, the two-mode approximation that one makes by projecting 
the problem on the eigenmodes 
$u_0$ and $u_1$ fails as the contribution of the other eigenstates cannot be neglected anymore.
 
Furthermore, another intuitive argument can be provided to explain the disagreement between predicted and numerical results observed 
with lowering $N_1$: as was already said, the effective potential $V_\text{eff}(x,y) = V(x,y) + |\psi_1(x,y)|^2$ has been calculated 
deriving the term $|\psi_1(x,y)|^2$
from the density profile of the 
dark component of the VB dipole state 
with $N_2 = 0$ and is considered to be fixed in the whole calculation. 
Our model looks for the bifurcation to occur at a finite value of $N_2$, but neglects that, for this value, in the actual physical system,  
the density profile for the dark component is different from the one with $N_2 = 0$. Thus, 
in doing this approximation, we do not take into account the effect of the second component on the first due to the interaction between them.  
Considering $V_\text{eff}$ independent of $N_2$ is expected to be valid if $N_1$ is large enough, 
as the dark component will then have a robust configuration with respect to variations due to the intercomponent interaction. But, decreasing 
$N_1$, varying $N_2$ will start to influence both components in a sensitive way and, in particular, 
an evident variation of 
the density profile of the dark component is expected. Therefore, for low $N_1$, the assumption that $|\psi_1(x,y)|^2$ is independent of $N_2$ 
is no longer appropriate and the model described above cannot be expected to be valid anymore. Therefore, in Fig.~\ref{fig:vortexbright5}(a), 
values of $N_1$ smaller than $100$ are not taken into account.

Let us now come back to the expected bifurcation from the anti-symmetric
branch, once its destabilization occurs, towards an asymmetric VB 
dipole solution.
Indeed, an example of this bifurcation has been illustrated in Fig.~\ref{fig:vortexbright5}(b).
Note that this diagram is the direct analog of  Fig.~4 in \cite{symmetry}, where symmetry-breaking bifurcations in a one-dimensional static double well were studied. 
Fig.~\ref{fig:vortexbright5}(c) illustrates a prototypical example
of the daughter state in the form of an asymmetric vortex-bright
dipole. 

The robustness of an asymmetric VB dipole of the type shown in Fig.~\ref{fig:vortexbright5}(c), 
combined with the Hamiltonian dynamics
of the system, is what gives rise to the tunneling observations
of Fig.~\ref{fig:vortexbright3}(b).
Naturally, there are two such asymmetric states, 
depending on which vortex-induced well picks up part of the bright
mass of the other, justifying the pitchfork character of the 
bifurcation. 
The BdG spectrum of the asymmetric state (not shown
here) reveals
the complete stability of the latter, in accordance with the
expectations from the above bifurcation theoretic arguments. 

\begin{figure}[ht]
 \centering
 \includegraphics[width = 0.4 \textwidth]{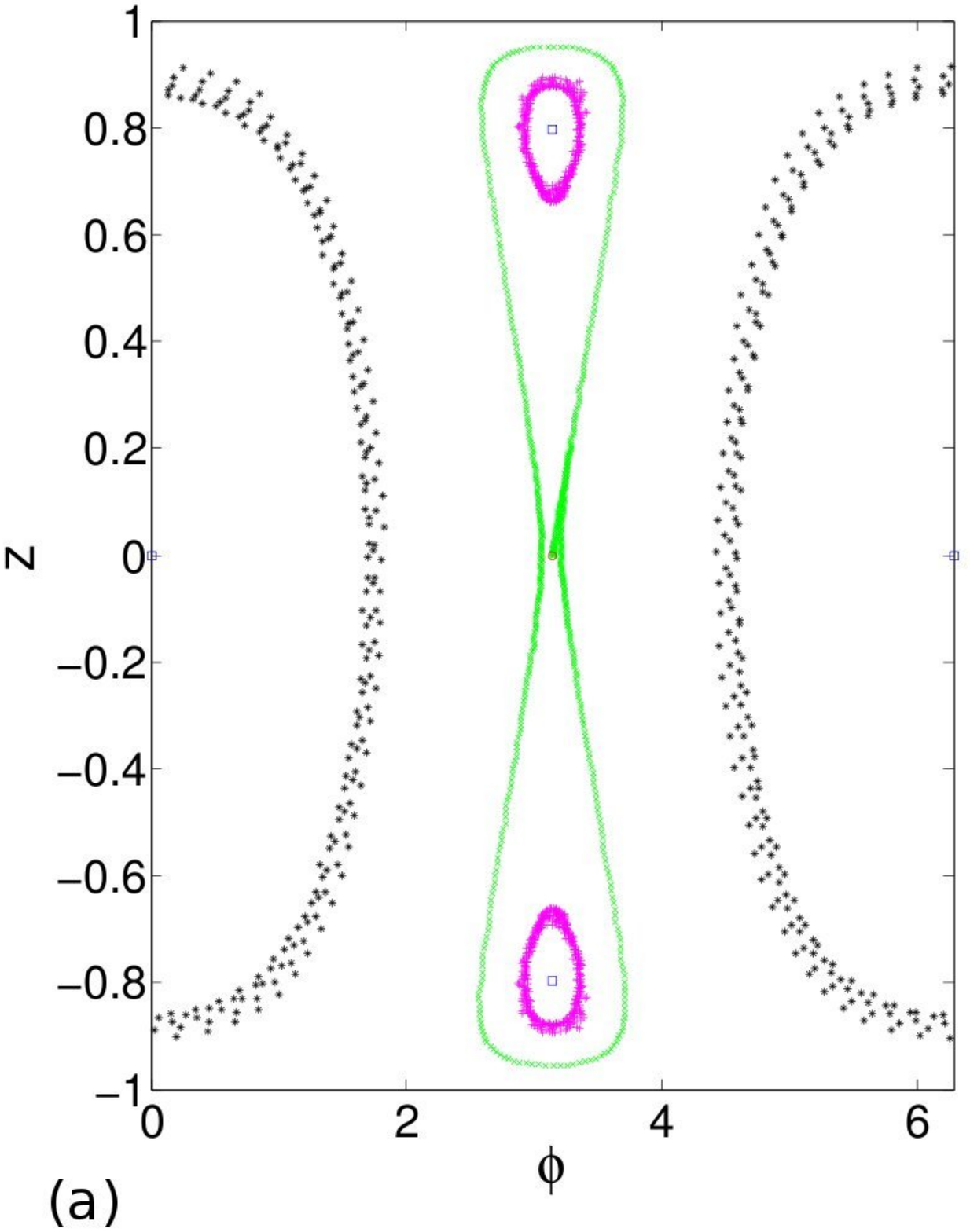}
\includegraphics[width = 0.4 \textwidth]{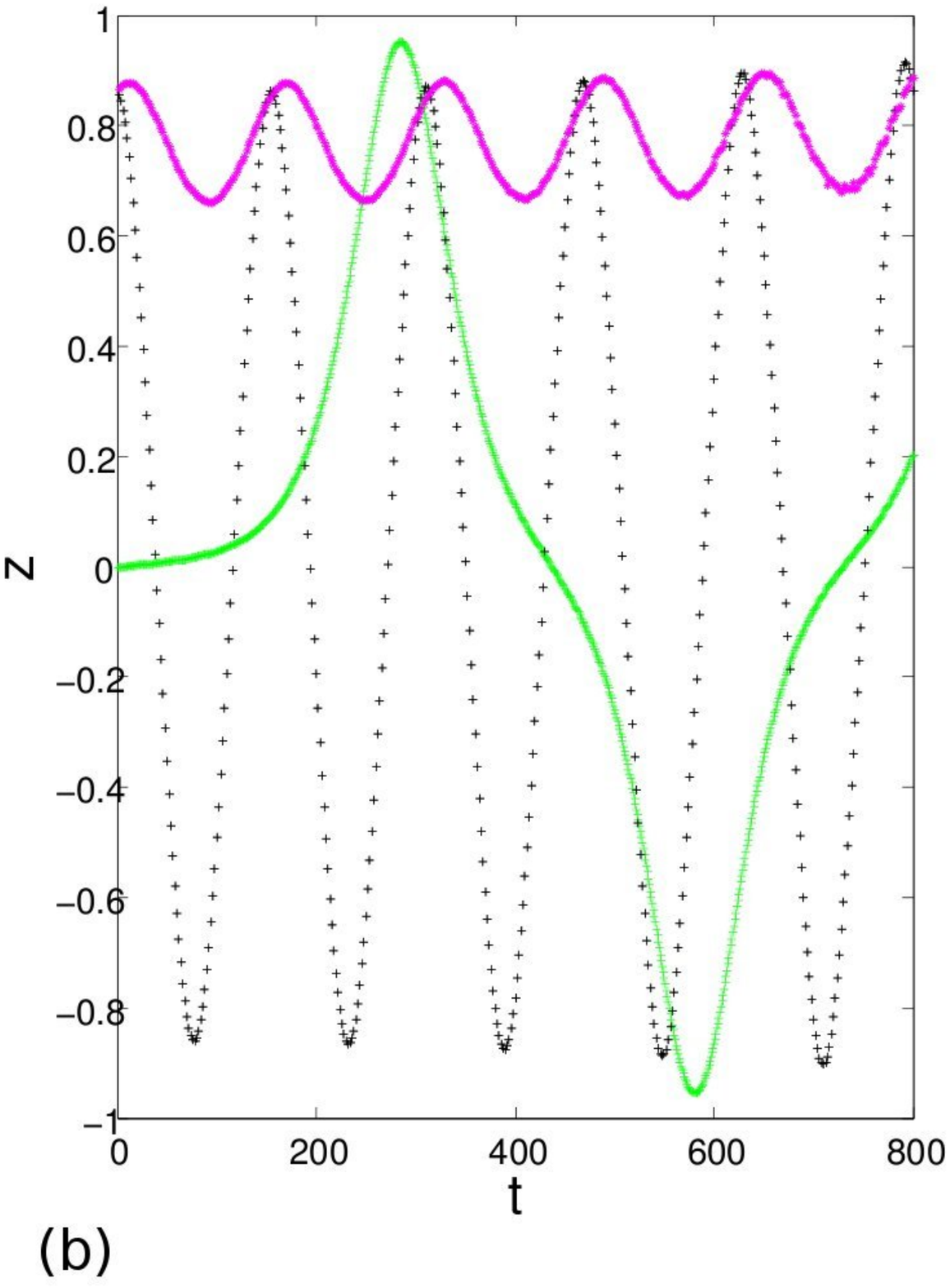}
\caption{(Color online) (a) The three types of trajectories of the bright component in the vortex-induced double well at $N_1 = 220$, $N_2 = 1$. 
The conjugate variables $z$, $\phi$ denote the population imbalance and phase difference between the two wells, respectively. 
All three types of expected trajectories can be seen to exist: oscillations around the asymmetric, stable fixed points (magenta, $\ast$ markers), 
plasma oscillations between the wells around an average phase of 0 (black, $+$ markers) and $\pi$ oscillations, i.e. tunneling oscillations 
with an average phase difference of $\pi$ between the two vortex cores (green, $\times$ markers, same data as in Fig. \ref{fig:vortexbright3}). 
The plot also shows the in-phase, out-of-phase and asymmetric fixed points of the system. (b) $z(t)$ dependence for the three trajectories, presented in the same color-markers code.}
\label{fig:phasespace}
\end{figure}

\begin{figure}[h!]
 \centering
  \fbox{\includegraphics[width = .9 \textwidth]{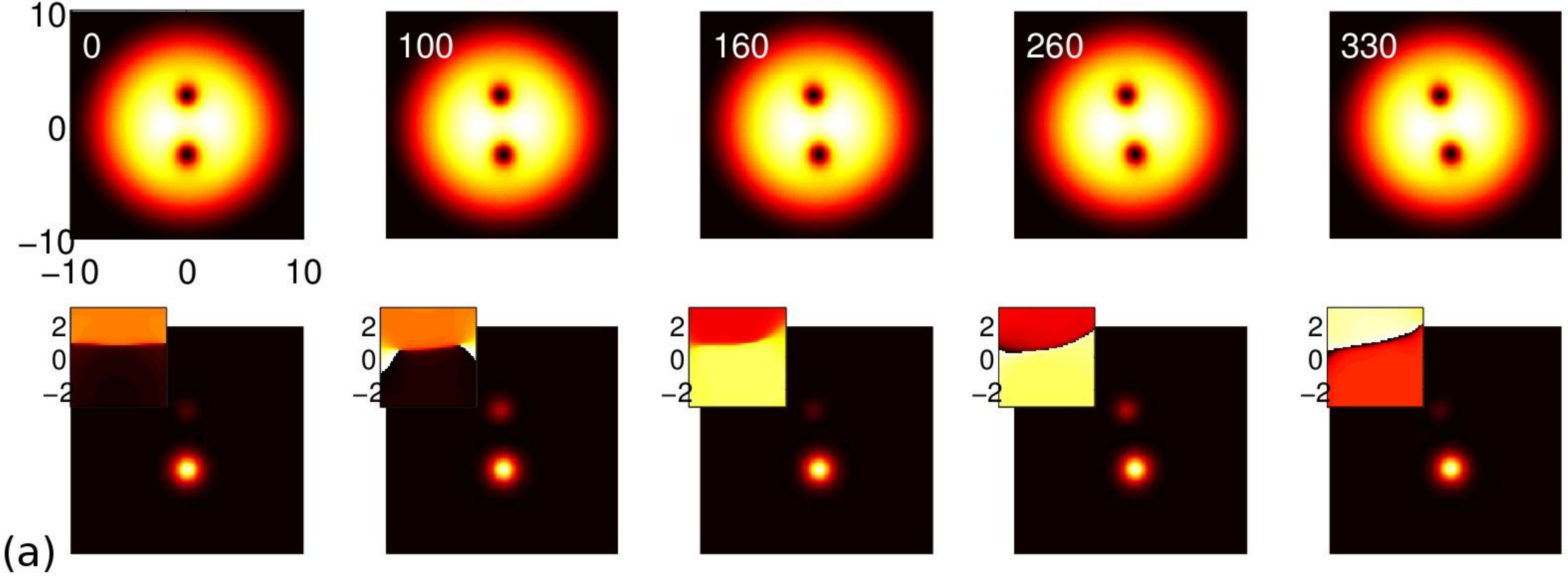}}
\fbox{\includegraphics[width = .9 \textwidth]{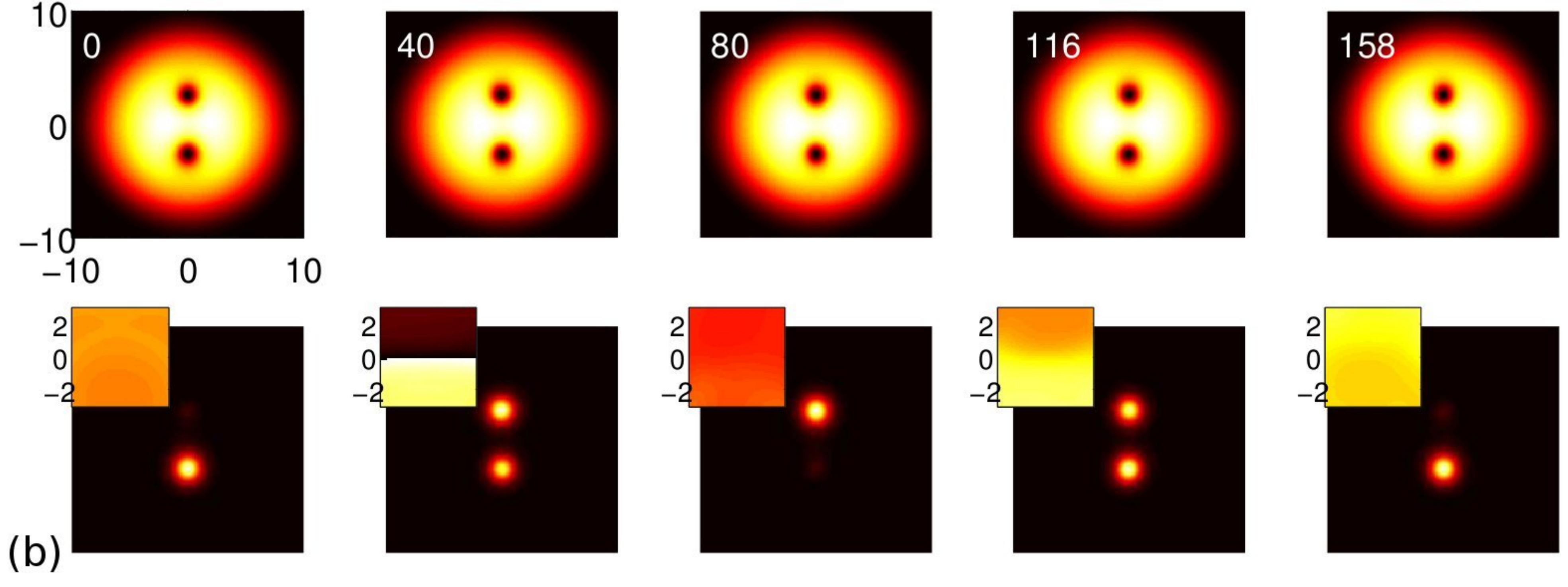}}
 \caption{(Color online) Simulated density profiles of both components and phase profiles of the bright component during an oscillation around one of the self-trapped states (a) 
and during a plasma oscillation (b), same data as included in the phase portrait Fig. \ref{fig:phasespace}.}
 \label{fig:vortexbrightnewcorner}
 \end{figure}

Let us in the following further explore the analogy between the bright component effectively trapped by the dark vortex dipole and a single species BEC in an external double well potential.
The possible types of dynamics in such a double well setting (sometimes called a {\it bosonic Josephson junction}) have attracted a lot of attention and have been studied both theoretically and experimentally in great detail, 
see e.g.~\cite{oberthaler} and references therein. The existence of asymmetric (``self-trapped'') solutions in such a setting was first studied in \cite{BJJ1,BJJ}.
We have performed a number of numerical simulations to show that starting from suitable initial conditions for the bright component in the VB dipole, all types of characteristic dynamics in such a bosonic Josephson junction can be recovered.
In particular, in the parameter regime where asymmetric states have bifurcated from the antisymmetric solution, three profoundly different types of dynamics are possible, namely: oscillations around the self-trapped, 
asymmetric configurations, tunneling oscillations between the two wells with a mean relative phase of 0 (``plasma oscillations'' around the in-phase fixed point) 
and such with a mean relative phase of $\pi$, the so-called ``$\pi$ oscillations'' around the $z$-symmetric out-of-phase fixed point, see e.g. \cite{oberthaler}.
These three types of trajectories are most easily distinguished in phase space of the conjugate variables $(z, \phi)$, where $z$ denotes the population imbalance between the two wells and $\phi$ is the corresponding phase difference.
More formally, these two quantities can be extracted from the full wavefunction within the two-mode approximation introduced above, taking into account only the ground and first excited states of the double well.
Starting from the known lowest symmetric and antisymmetric eigenfunctions $u_0$ and $u_1$, modes which are localized in one of the wells can be formed, namely 
$u_{D,U} = (u_0 \pm u_1)/\sqrt{2}$.
Expanding the full wavefunction in this basis, $\psi_2 = \sqrt{\rho_U} \exp(i \phi_U) u_U + \sqrt{\rho_D} \exp(i \phi_D) u_D$, where both $\rho_{U,D}$ and $\phi_{U,D}$ are real, 
the phase space variables are calculated as $\phi = \phi_U - \phi_D$, $z = (\rho_U - \rho_D)/(\rho_U + \rho_D)$. 
A number of different trajectories of the bright component in the vortex-induced double well, presented in terms of these $(z, \phi)$ variables, 
are shown in Fig.~\ref{fig:phasespace}(a) for parameter values $N_1 = 220$, $N_2 = 1$. In Fig.~\ref{fig:phasespace}(b), 
the $z(t)$ dependence for three different trajectories is shown.
There is full agreement of this phase portrait with the one obtained for a single species bosonic Josephson junction e.g. in \cite{oberthaler}, indicating that for the small populations in the bright component we are considering here, the approximation of ``freezing'' the dark component into an effective double well potential is a very good one.
Finally, Fig.~\ref{fig:vortexbrightnewcorner} collects density and phase profiles of the two remaining trajectories included in the phase portrait, namely the one revolving around one of the self-trapped, asymmetric states (Fig.~\ref{fig:vortexbrightnewcorner}(a)) 
and a ``plasma oscillation'' (Fig.~\ref{fig:vortexbrightnewcorner}(b)) around the symmetric, in-phase configuration (while the ``$\pi$ oscillation'' around 
the antisymmetric fixed point was already shown in Fig.~\ref{fig:vortexbright3}(b)).

\begin{figure}[h!]
 \centering
 \includegraphics[width = 1 \textwidth]{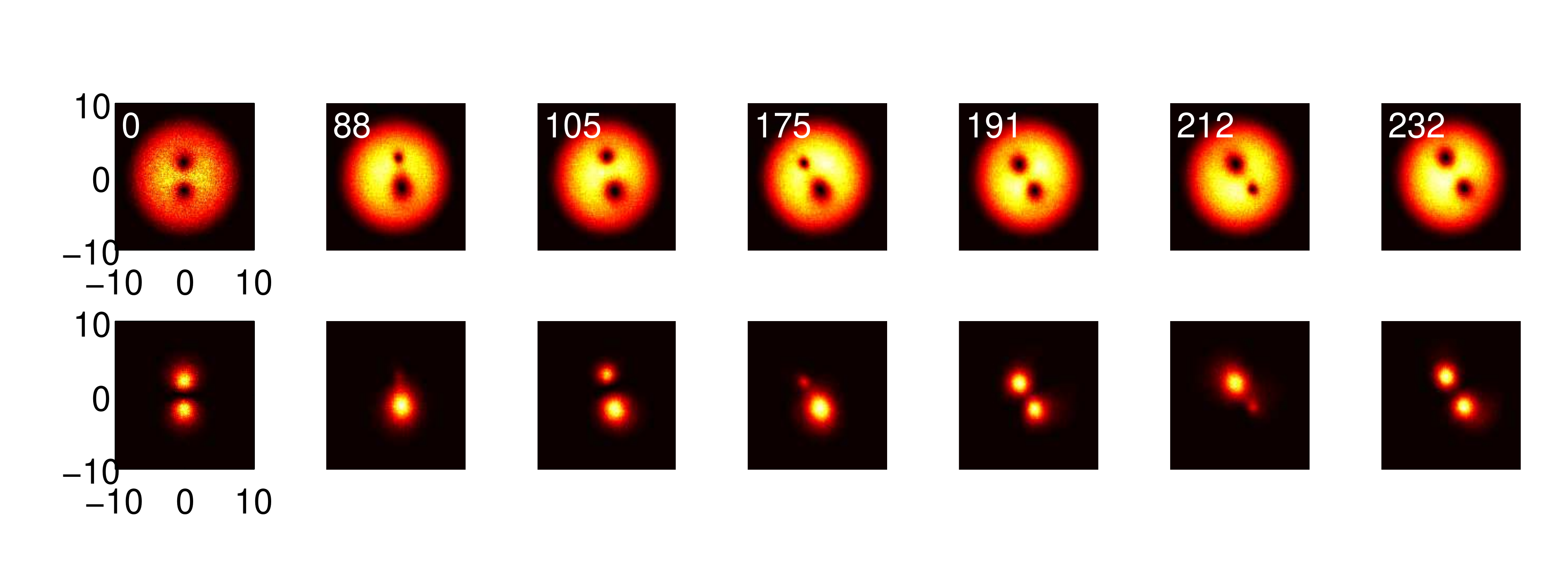}
 \caption{(Color online) Time propagation of an out-of-phase vortex-bright soliton dipole, perturbed with white noise at $N_1 = 390$ and $N_2 =  50$.}
 \label{fig:vortexbright4}
 \end{figure}

So far, we have concentrated this section's discussion of the tunneling dynamics on the regime of small $N_2$, 
close to the linear limit of the effective Hamiltonian introduced by assuming the dark component to be frozen.
While this is essential for the simplified semi-analytical model introduced above to make the connection to symmetry-breaking in a double well potential, 
we have found direct numerical evidence that the out-of-phase VB dipole is unstable towards tunneling dynamics in the bright component
even in parameter regimes where $N_2$ is considerably larger, see Fig.~\ref{fig:vortexbright4} for an example.
In this case the larger population of the bright component now leads to a stronger back-coupling to the dark component. 
In particular, it can now be clearly observed that the vortex cores forming the dipole in the dark component change size following 
the trend of the bright solitons filling them. 
While in Fig.~\ref{fig:vortexbright3}(b) the tunneling oscillations, past the initial transient of asymmetry
build-up, directly turn periodic, this is no longer the case here. 
Instead, in the run shown in Fig.~\ref{fig:vortexbright4},  the bright component first almost completely tunnels to the lower well, but then oscillates between this asymmetric and a more symmetric occupation 
of the two wells (see the first four timesteps shown). Only after that, a majority of bright component atoms enters the upper well (second to last timestep), 
and then directly oscillates back to approximately equal occupation of the vortex cores (last timestep). The conclusion of 
this first period is followed by a series of nearly equal period similar oscillation steps.

Another relevant difference to Fig.~\ref{fig:vortexbright3}(b) is that the tunneling dynamics is accompanied 
by a rotational motion of the vortices, while for smaller $N_2$ the dipole essentially stayed aligned along 
the trap's axis during the propagation. 
This may be taken as a hint that for a larger $N_2/N_1$ ratio the interspecies interaction may induce a coupling between
the unstable tunneling-type mode of the bright component and the precessional degrees of freedom of the vortices in the dark component.

\section{Conclusions and Future Challenges}

In the present work, we have revisited the vortex-bright solitary wave
states and have illustrated their prototypical generalization to 
vortex-bright soliton dipoles. This is a first step towards the realization
of two-component clusters of such entities. In the process, we have
explored a number of interesting features. Firstly, even a single
vortex-bright soliton was found to exhibit a transition from a saddle point in 
the energy to a local minimum, as its bright component becomes
more significant. Secondly, two types of VB dipoles were identified,
in analogy with their dark-bright one-dimensional counterparts.
The first of them had the bright pulses (trapped 
inside the vortex cores) be in phase, while the other one had them
as out of phase. We have identified the bifurcation of these
states, as stemming from the corresponding dark-bright stripe
states, one of which has the bright second component in phase
(stemming from the ground symmetric state) and one of which has it out
of phase (stemming from the first excited anti-symmetric state).
Subsequently, the dynamical stability of the daughter VB dipole
states emerging from these bifurcations was quantified. It was
found that the in-phase one is generically stable, while the
out of phase structure suffers an exponential instability (past
a relevant critical point).
The latter gave rise to tunneling dynamics and spontaneous symmetry
breaking manifestations. This phenomenology was understood on the
basis of an effective double well model, where the vortex cores played 
the role of the wells and led to asymmetric VB dipoles.

This study paves the way for a more detailed understanding 
of multi-component structures. One of the key items that
remain open concerns the effective description of such states.
In particular, it is remarkable that these states have both
a solitonic character through their bright component and a vortex
character through their dark one. It is then particularly interesting
to examine how the effective equations characterizing the interactions
of such entities look and what the corresponding interplay is between
the vortex and the solitonic character. Another natural direction
concerns generalizing the ideas presented herein to a setting with
more vortex-bright dipoles e.g. three such, which might naturally
be observable in two-component generalizations of the experiments of~\cite{bagnato}.
In fact, careful inspection of the figures presented herein (see e.g.
the second pair of imaginary eigenfrequencies stemming from the stripe
states' BdG analysis in the bottom panel of Fig.~~\ref{fig:vortexbright2})
already suggests the bifurcation of such states.
Furthermore, extending the present considerations to three-dimensions
and towards a more detailed understanding of bound states of
vortex rings~\cite{pantoflas} 
(and two-component generalization thereof) would
also constitute an important theme for future explorations. Such studies
are presently in progress and will be reported in future works.

\end{document}